\documentclass[twocolumn,amsmath,amssymb,floatfix,nofootinbib,prc]{revtex4}
\usepackage{graphicx}
\usepackage{bm}
\usepackage{amsmath}
\usepackage{epsfig}

\begin{document}

\title{Morphology of High-Multiplicity Events in Heavy Ion Collisions}
\author{P. Naselsky, C. H. Christensen, P. R. Christensen, P. H. Damgaard, A. Frejsel, J. J. Gaardh\o 
je, A. Hansen, M. Hansen, J. Kim}
\affiliation{Discovery Center, Niels Bohr Institute, Blegdamsvej 17, DK-2100 Copenhagen, Denmark}
\author{O. Verkhodanov}
\affiliation{Special Astrophysical Observatory, Russian Academy of Sciences, Nizhnij Arkhyz, Russia}
\author{U.A. Wiedemann}
\affiliation{Physics Department, Theory Unit, CERN, CH-1211 Geneve, 23, Switzerland}

\begin{abstract}
We discuss opportunities that may arise from subjecting high-multiplicity events in relativistic heavy ion collisions to an analysis 
similar to the one used in cosmology for the study of fluctuations of the Cosmic Microwave Background (CMB). To this end, we 
discuss examples of how pertinent features of heavy ion collisions including global characteristics, signatures of collective flow and event-wise 
fluctuations are visually represented in a Mollweide projection commonly used in CMB analysis, and how they are statistically analyzed 
in an expansion over spherical harmonic functions. 
If applied to the characterization of purely azimuthal dependent phenomena such as collective flow, the expansion coefficients of spherical 
harmonics are seen to contain redundancies compared to the set of harmonic flow coefficients commonly used in heavy ion collisions. 
Our exploratory study indicates, however, that these redundancies may offer novel opportunities 
for a detailed characterization of those event-wise fluctuations that remain after subtraction of
 the dominant collective flow signatures. 
 By construction, the proposed approach allows also for the characterization of more complex 
  collective phenomena like higher-order flow and other sources of fluctuations, and it may be extended to the 
  characterization of  phenomena of non-collective origin such as jets. 
\end{abstract}

\preprint{CERN-PH-TH/2012-74}
\maketitle

\section{Introduction \label{sec:intro}}

The bulk of low-momentum particles produced in heavy ion collisions at the Relativistic Heavy Ion Collider (RHIC) and at the Large Hadronic Collider (LHC)
appears to emerge from a common flow field that is
largely driven by the density gradients produced in the earliest stage of the collision~\cite{Heinz:2009xj,Romatschke:2009im,Teaney:2009qa,Hirano:2007gc}. 
This fluid dynamic paradigm of heavy ion collisions is supported by a large set of data from the LHC~\cite{ALICE-flow-PbPb,Aad:2012bu,Velkovska:2011zz} and 
RHIC~\cite{BRAHMS-WP, PHOBOS-WP, PHENIX-WP, STAR-WP, STAR-flow2001, PHENIX-flow2003} 
which includes the dependence of particle production on the azimuthal angle relative to the reaction plane and its dependence on transverse momentum,pseudorapidity $\eta$,
 particle species and centrality (impact parameter) of the collision. 
Complementary support
comes from recent indications that also event-by-event fluctuations of the energy density distribution
produced in the earliest stage of a heavy ion collision, are propagated fluid dynamically. In 
particular, the measurement of non-vanishing odd harmonic coefficients, $v_{n}$, in the 
azimuthal dependence of particle distributions is an unambiguous signal for the presence
of sizable fluctuations in the initial stage of a heavy ion collision \cite{Alver-Roland-2010,Teaney:2010vd,Alver:2010dn,Sorensen2010,Bhalerao:2011yg,PHENIX-vn-flow, ALICE-vn-flow}. 

These findings point to a remarkable set of commonalities between the 
physics of the "Big Bang" determining the evolution of the Universe, and the physics of the 
"Little Bangs" formed in heavy ion collisions, that creates hot and dense matter under conditions akin to those prevailing in the early universe around the first microsecond, where the transition from the Quark Gluon Plasma (QGP) to a confined hadron gas (HG) occurred. 

In both cases, the physical system is viewed at
an initial time as exhibiting a phase space distribution with a high degree of 
symmetry, overlaid with distributions of localized fluctuations. In both cases, these
initial fluctuations are propagated fluid dynamically and are experimentally accessible, 
in the first case as fluctuations of the (photon) Cosmic Microwave Background (CMB)~\cite{COBE, WMAP,PLANCK,BOOMERANG, MAXIMA, CBI, ACT} and,
 in the second case, as fluctuations and characteristic variations in the density of particles produced in very energetic heavy ion collisions~\cite{Mishra,Mocsy}. Also, in both cases, the decoupling of the different particle species from the common fluid dynamical system provides important possibilities for experimental verification of the collective dynamics. In the case of Big Bang physics, it sets the time scale for the
decoupling of photons and determines the primordial abundances of light nuclei. In the
case of heavy ion physics, it sets, i.a., the hierarchy of temperatures for chemical
and kinematic freeze-out, and determines the relative abundances of hadronic resonances. 
Furthermore, the study of fluctuations is motivated mainly by
the idea that details of the fluid dynamical evolution of a system make it possible to constrain its material properties. 
In the case of Big Bang physics, this has allowed to constrain the composition of the Universe in
terms of visible matter, dark matter and dark energy~\cite{WMAP}. In the case of heavy ion collisions, this has allowed to establish that the produced matter exhibits the properties
of an almost perfect fluid with a shear viscosity to entropy density ratio that is close to minimal~\cite{Teaney:2003kp,Romatschke2007,Heinz:2009xj,Romatschke:2009im,Teaney:2009qa,Hirano:2007gc}.

The question arises whether these qualitative analogies between the physics of the Early Universe
and the physics of heavy ion collisions can provide conceptual or technical advances in either field.
Can heavy ion physics learn from the techniques developed to analyze Big Bang physics, or vice versa, 
can it contribute to advances in modern cosmology? Synergies between both fields may arise on different levels. 
For instance, cosmology has studied in significant detail the question of how fluid dynamic perturbations 
are propagated in an expanding system, and there is at least some understanding of how
non-linearities and turbulent phenomena relevant for structure formation build up, and which experimental measures
may give access to them. In the context of heavy ion collisions, first analyses of the propagation of fluid dynamic 
perturbations have been undertaken with analytical and numerical 
techniques~\cite{Holopainen:2010gz,Qiu:2011hf,Schenke:2011bn,Gubser:2010ui,Staig:2010pn,Florchinger:2011qf}
, but these are arguably at the very beginning, 
and further explorations in close analogy to developments in cosmology may be beneficial. 

Similarities between the Big Bang and the Little Bangs may also lead to synergies on the level of the techniques for the analysis of data on fluctuations. 
The main purpose of the present paper is to initiate such an
effort. To this end, we shall use here a standard tool \cite{glesp} for analyzing the CMB anisotropy,
and apply it to the study of sets of simulated particle densities from ultra-relativistic heavy ion collisions. 

We note that studies of the physical system produced in the Big Bang and the systems produced in heavy ion 
collisions share not only remarkable commonalities but have also significant differences. For an analysis 
of fluctuations and their separation from other sources of signals, some of the obvious differences
(such as the vastly different scales involved in both problems, the different fundamental forces that
govern the dynamics, and the difference in formulating the dynamics in general relativity or special relativity) may
be less important. Potentially relevant, however, is the difference between a study of
one 'big' event (even if consisting of a large but finite sample of statistically independent patches) and the study of
an - in principle arbitrarily abundant - sample of mesoscopic events. 

The paper is organized as follows. In section 2 we introduce the spherical harmonic approach that underlies CMB data analysis.
In sections 3 and 4, we present and discuss an application of this approach to simulated heavy ion data,
and in section 5 we apply the method to the particular case of elliptic flow, a standard measurement in heavy ion collisions. 
 In section 6 we summarize our findings and we present a brief outlook. Finally, in Appendix A we have presented mathematical
details of the method and generalization of the model with constant flow amplitude to more complex one, when the amplitude of the flow depends on pseudorapidity .

%%%%%%%%%%%%%%%%%%%%%%%%%%%%%%%%%%%%%%%%%%%%%%%%%%%%%%%%%%%%%%%%%%%%%%%%%%%%%
\section{Formulation of the problem \label{sec:problem} }
 
In cosmology, the CMB anisotropy is analyzed
by means of two-dimensional maps that essentially cover the full sky 
(like those from the COBE, WMAP and PLANCK experiments \cite{COBE, WMAP,PLANCK}), or only patches 
of the sky (Boomerang, MAXIMA, CBI, ACT etc.\cite{BOOMERANG, MAXIMA, CBI, ACT}). Analysis of such maps reveal the small 
temperature fluctuations and anisotropies that the early Universe has left on top of an 
otherwise smooth and constant background. By expressing the measured signal in terms of 
an expansion in spherical harmonics, useful quantities such as the power spectrum, 
alignment between different multipoles, statistical anisotropy and Gaussianity of the CMB
can be constructed and compared with theory even from a single map (realization) of the CMB sky \cite{PLANCK}.
In the analysis of CMB data, the methods focus on the separation
of the different components of the observed signal from that of the primordial CMB. The by-product of this separation yields additional maps for different kinds of foreground effects (e.g. synchrotron, free-free and dust emission) and the point-sources catalog \cite{PLANCK1}.

The analysis of heavy ion data from nuclear collisions at the LHC and the RHIC faces similar challenges. 
On one hand, several independent statistical methods are in use~\cite{Ollitrault1992,VPS-review2010} 
to analyze two- and multi-particle 
correlations in terms of flow coefficient $v_{n}$, taken to be sensitive to the fluid dynamical properties of
 the high density partonic system in its early stages of expansion. On the other hand, the same
particle correlations are known to be sensitive to other important dynamical features of heavy ion collisions,
such as jets or resonance decays. Typical tasks in heavy ion collisions are then to either clean the flow signal 
as far as possible from confounding dynamical features, or to analyze a different class of interesting physical 
phenomena (such as the 'point-sources' produced by jets) without biasing its analysis by underlying collective effects. 
The question arises, what one could learn for this class of problems if the investigation of morphological 
features commonly done in the CMB data analysis was applied directly to heavy ion collisions?

Each heavy ion collision results in a distribution $f(\eta,\phi)$ of particles as a function of 
azimuthal angle $\phi$ and pseudorapidity $\eta$. Pseudorapidity is related to the polar angle $\theta$ of 
the produced particle with respect to the beam direction via $\eta=-\ln(\tan(\frac{\theta}{2}))$, and $\phi$ characterizes the angle in the transverse
plane orthogonal to the beam. We simulate such distributions with HIJING \cite{HIJING} (Heavy-Ion Jet INteraction Generator), 
which is an event generator that reproduces a number of main features of heavy ion physics in the energy range of RHIC and LHC. 
It is build on PYTHIA \cite{Pythia64} routines for hard interactions
and JETSET \cite{JETSET-Sjostrand1986} routines for string fragmentation. The initial geometry and matter distribution is 
determined via a Glauber model \cite{Glauber,Bialas:1976ed}. Various phenomenologically relevant nuclear effects are also 
implemented, including e.g. the nuclear modification of parton distribution functions, or parametrization of
the expected parton energy loss. However, HIJING does not directly model a collective dynamics resulting in flow.
As this is relevant for our study, we include flow effects a posteriori by modulating the azimuthal distribution of 
particles from HIJING.

%%%%%%%%%%%%%%%%%%%%%%%%%%%%%%%%%%%%%%%%%%%%%%%%%%%
\begin{figure}[!b]
 \begin{center}
 \centerline{\includegraphics[scale=0.52]{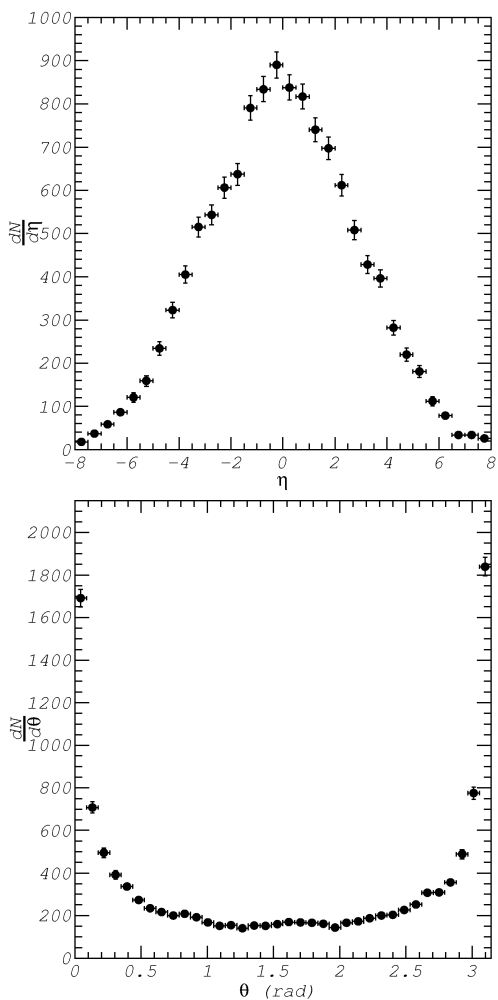}}
  \caption{Top: Pseudorapidity density of charged particles simulated by HIJING for a single semi-central collision between two Pb nuclei at current LHC energies ($ \sqrt{s_{nn}}=2.76 TeV $). Bottom: The same distribution as a function of the polar angle $\theta$ relative to the beam direction.}
  \label{fig_dNdEta}
 \end{center}
\end{figure}
%%%%%%%%%%%%%%%%%%%%%%%%%%%%%%%%%%%%%%%%%%%%%%%%%%%

Particle production in heavy ion collisions is known to display a regular $\eta$-dependence that peaks around $ \eta=0 $ and exhibits a forward-backwards symmetry 
in the case of collisions between identical nuclei. Fig.~\ref{fig_dNdEta} shows results for a semi-peripheral Pb-Pb collision at the LHC with a total of 12316 particles produced 
over more than 16 units of pseudo-rapidity. The pseudo-rapidity distribution of this randomly chosen simulated single event
shows visible but relatively small fluctuations 
around a smooth event-averaged distribution. In preparation of the following analysis, we have replotted this distribution as a function of the polar angle $\theta$.
We see that the $\theta$-distribution is approximately flat over an extended range $\frac{\pi}{4} < \theta < \frac{3\pi}{4}$  which corresponds to a sizeable window
around mid-rapidity ($-0.88 < \eta < 0.88$). A presentation in the variable $\theta$ thus highlights the region of approximately one unit around mid-rapidity for which 
most differential data are available. 

\begin{figure}[!b]
 \begin{center}
  \centerline{\includegraphics[scale=0.7]{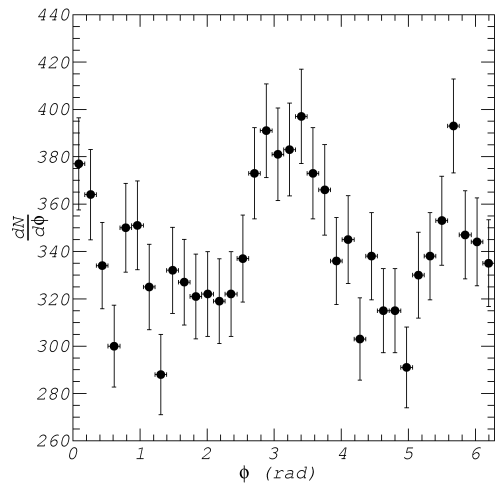}}
  \caption{Density of charged particles as a function of the azimuthal angle $\phi$ in the plane transverse to the beam direction, simulated by HIJING and
  modified by an elliptic flow with $v_{2}=0.10$.} 
  \label{fig_dNdPhi}
 \end{center}
\end{figure}

As remarked already, HIJING does not generate realistic flow signals. For our study, we superimpose such flow signals a posteriori on HIJING events
by redistributing the generated particles  such that event-averaged ensembles  of collisions show a second order harmonic oscillation of the azimuthal angle $\phi$ 
with some prescribed magnitude $v_2$. This procedure results in events that carry azimuthal oscillations 
supplemented by event-by-event fluctuations. Fig.~\ref{fig_dNdPhi} shows a typical event realization 
for an average flow amplitude $v_2 = 0.10$. We caution the reader that experimental data on flow reveal characteristic dependencies  of $v_2$ on transverse momentum, 
rapidity and particle identity that are not included in this ad hoc simulation. The simplified procedure adopted here will not account for all phenomenologically relevant
properties of particle flow, but it will be sufficient for the purpose of our study.

%%%%%%%%%%%%%%%%%%%%%%%%%%%%%%%%%%%%%%%%%%%%%%%%%%%
%%%%%%%%%%%%%%%%%%%%%%%%%%%%%%%%%%%%%%%%%%%%%%%%%%%

\section{Spherical harmonic decomposition and analysis of the symmetries of the model \label{sec:decomp} }

The harmonic analysis of particle production in heavy ion collisions focus typically on  
the Fourier decomposition of the $\phi$-dependence (see Fig. \ref{fig_dNdPhi}) over a narrow region in pseudorapidity around $\eta=0$. Here, we contrast this procedure with the one employed in the CMB data analysis that starts from an expansion of the entire two-dimensional map $f(\theta,\phi)$ in terms of 
spherical harmonics $Y_{l,m}(\theta, \phi)$,
\begin{eqnarray}
&&	f(\theta,\phi) = \sum_{l=1}^{l_{max}}\sum_{m=-l}^{l} a_{l,m}Y_{l,m}(\theta, \phi), \nonumber\\
&& Y_{l,m}(\theta,\phi)=\sqrt{\frac{(2l+1)}{4\pi}\frac{(l-m)!}{(l+m)!}}P^m_l(x)e^{im\phi}=\nonumber \\
&&=N_{l,m}P^m_l(x)e^{im\phi}\, 
	\label{decomposition}
\end{eqnarray}
where $ a_{l,m}=|a_{l,m}|\exp(-im\Phi_{l,m})$ are the coefficients of decomposition with amplitudes
$|a_{l,m}|$ and phases $\Phi_{l,m}$ for each component $l,m$, $P^m_l(\cos\theta)$ are the associated Legendre polynomials, and 
\begin{eqnarray}
 C^S(l) =\frac{1}{2l+1}\sum_{m=-l}^l|a_{l,m}|^2
\label{pow}
 \end{eqnarray}
is the total power spectrum.

%%%%%%%%%%%%%%%%%%%%%%%%%%%%%%%%%%%%%%%%%%%%%%%%%%%%%%%%
\begin{figure}[!b]
 \begin{center}
  \centerline{\includegraphics[scale=0.37]{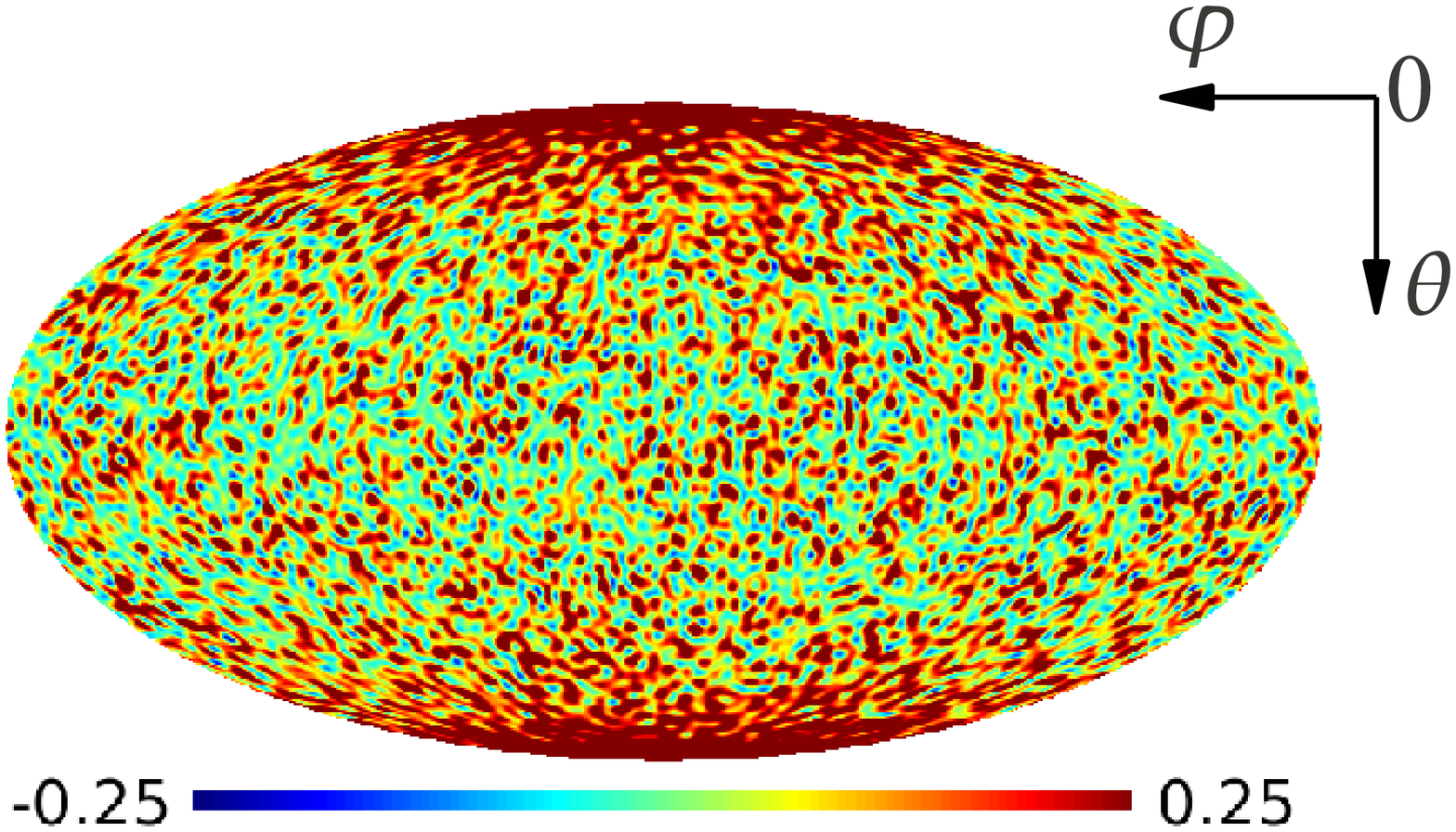}}
	\centerline{\includegraphics[scale=0.34]{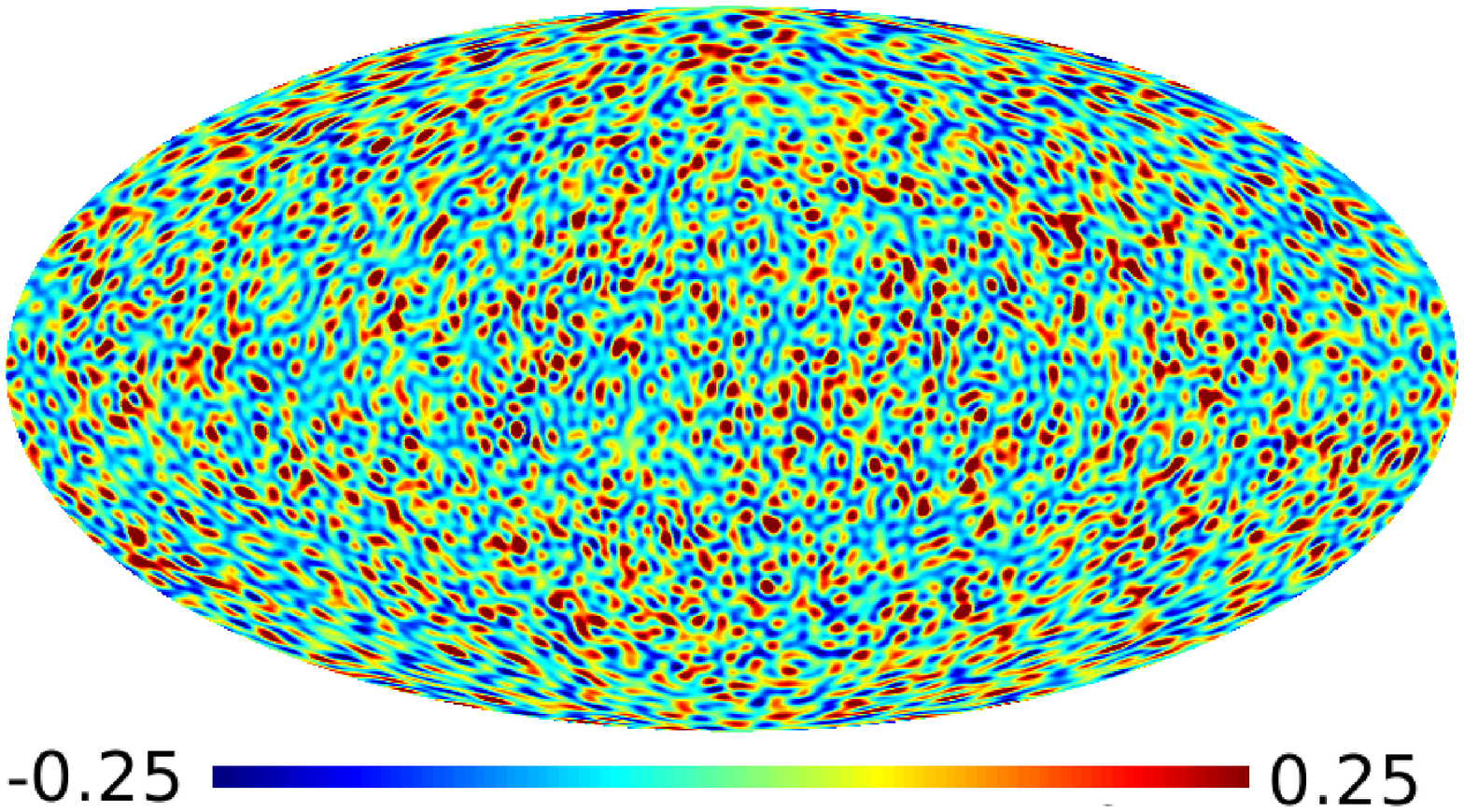}}
  \caption{(Color online) Top: Mollweide projection of a single HIJING event without flow. The azimuthal angle $\phi$ runs along the 'equator', the polar
  angle $\theta$ runs from pole to pole.
  The color coding tracks deviations relative to the average level of counts. Bottom: The $m=0$ mode has been removed from the map. 
The angular resolution of the maps corresponds to $l_{max} = 100$ (see Eq. (\ref{decomposition}). The multiplicity of the event is $17000$.}
  \label{fig_Mollweide_map}
 \end{center}
\end{figure}
%%%%%%%%%%%%%%%%%%%%%%%%%%%%%%%%%%%%%%%%%%%%%%%%%%%%%%%%%
As it is seen from Eq.~(\ref{decomposition}), the constant term with $l=0$ which corresponds to the average density of
the particles on the sphere is not included in our analysis.  Thus, $f(\theta,\phi)$ represents the fluctuations of the
number density of particles with respect to the mean value and it can take positive or negative values. 
The parameter $l_{max}$ in Eq.~(\ref{decomposition}) characterizes
the angular resolution of the map. In principle, infinitely precise pointing resolution requires  ($l_{max}\rightarrow\infty$) which is
computationally prohibitive. In the CMB analysis, one choses the parameter $l_{max}$ according to the angular resolution of the 
experimental data which is dictated by the segmentation of the detector.  For instance, $l_{max}=100$ corresponds to an angular 
resolution $\Delta=\pi/l_{max}\simeq 1.8^o$.In heavy ion physics, other physics considerations such as the typical angular 
separation of nearest charged tracks in the $\Delta \eta$ and $\Delta \phi$ may motivate the choice of $l_{\rm max}$. 
In the following, we use a resolution $l_{max}=100$ that is sufficiently high to reveal fluctuations on small scale.
We work for illustrative purposes with the standard GLESP \cite{glesp} pixelization where $l_{max}=100$ corresponds
to a number of pixels in the $\theta$-direction $N_{\theta}=2l_{max}+1$ and in the azimuthal direction $N_{\phi}=2N_{\theta}$.

\subsection{Mollweide projection of heavy ion events}

To familiarize ourselves with the main features of a CMB-like harmonic analysis of heavy ion collisions,
 we start from a 'baseline'
distribution $f(\theta,\phi)$ of a heavy ion collision in which features of global collective dynamics are absent. 
To this end, we consider as baseline
a HIJING event without superimposed flow signal ($v_2 = 0$). Fig.~\ref{fig_Mollweide_map} shows 
the corresponding fluctuations of the
distribution $f(\theta,\phi)$ in the so-called Mollweide projection that is heavily used in CMB analysis. 
In this representation the polar angle $ \theta $ runs vertically from the 'north pole' to the 'south pole' 
and the azimuthal angle $ \phi $ runs along the 'equator' 
from the center to  left . Let us pause to discuss in more detail the information shown in this map. 

We have seen already in the context of Fig.~\ref{fig_dNdEta} that particle
distributions are approximately flat in a wide window of the polar angle $\frac{\pi}{4} < \theta < \frac{3\pi}{4}$. Accordingly, the particle density and fluctuations 
in $\theta$ remain approximately unchanged in a broad band around the equator of the Mollweide projection (see upper part of Fig.~\ref{fig_Mollweide_map}).
Larger  particle densities are found at the poles reflecting the large number of particles emitted at small angles with respect to the beam direction (see bottom panel of Fig. \ref{fig_dNdEta}).  This enhanced particle yield at the poles is a kinematically trivial but potentially confounding factor for studies that aim at 
establishing event-by-event global or local signals on top of an
event-average background that varies significantly in $\eta$ or $\theta$. Thus, irrespective of the choice of coordinates, the question arises 
to what extend 
potentially interesting information about fluctuations and azimuthal asymmetries can be disentangled from the strongly varying background in $\eta$ or $\theta$.
The approximately uniform distribution of fluctuation in the lower panel of Fig.~\ref{fig_Mollweide_map} demonstrates that removing from the expansion of $f(\theta,\phi)$
the $m=0$ mode of the harmonic decomposition goes a long way towards achieving this aim. We note that the $m=0$ spherical harmonic components do not
depend on $\phi$ but only on $\theta$. Therefore, removing this component will remove the dominant dependence on the average pseudorapidity particle distribution
without modifying any flow signal.  In general, a dependence of particle distributions on pseudo-rapidity will remain after subtraction of the dominant $m=0$ component,
of course. In appendix A, we explain how an arbitrary $\eta$-dependence can be dealt with in the present formalism. For the sake of a particularly transparent 
presentation, we focus in the main text on the simpler case in which the $\eta$-dependence can be treated satisfactorily by subtracting the $m=0$ mode.

%%%%%%%%%%%%%%%%%%%%%%%%%%%%%%%%%%%%%%%%%%%%%%%%%%%%%%%%%
\begin{figure}[!h]
   \includegraphics[scale=0.72]{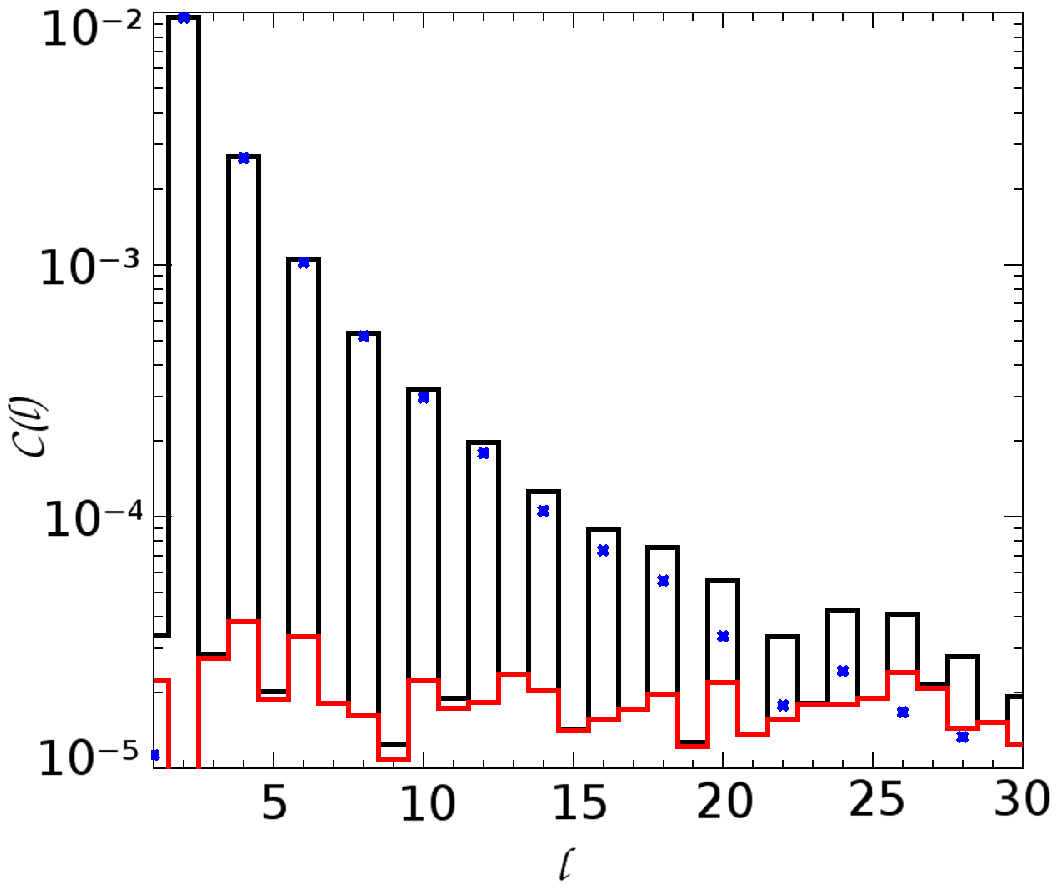}\vspace{0.5cm}
   \includegraphics[scale=0.25]{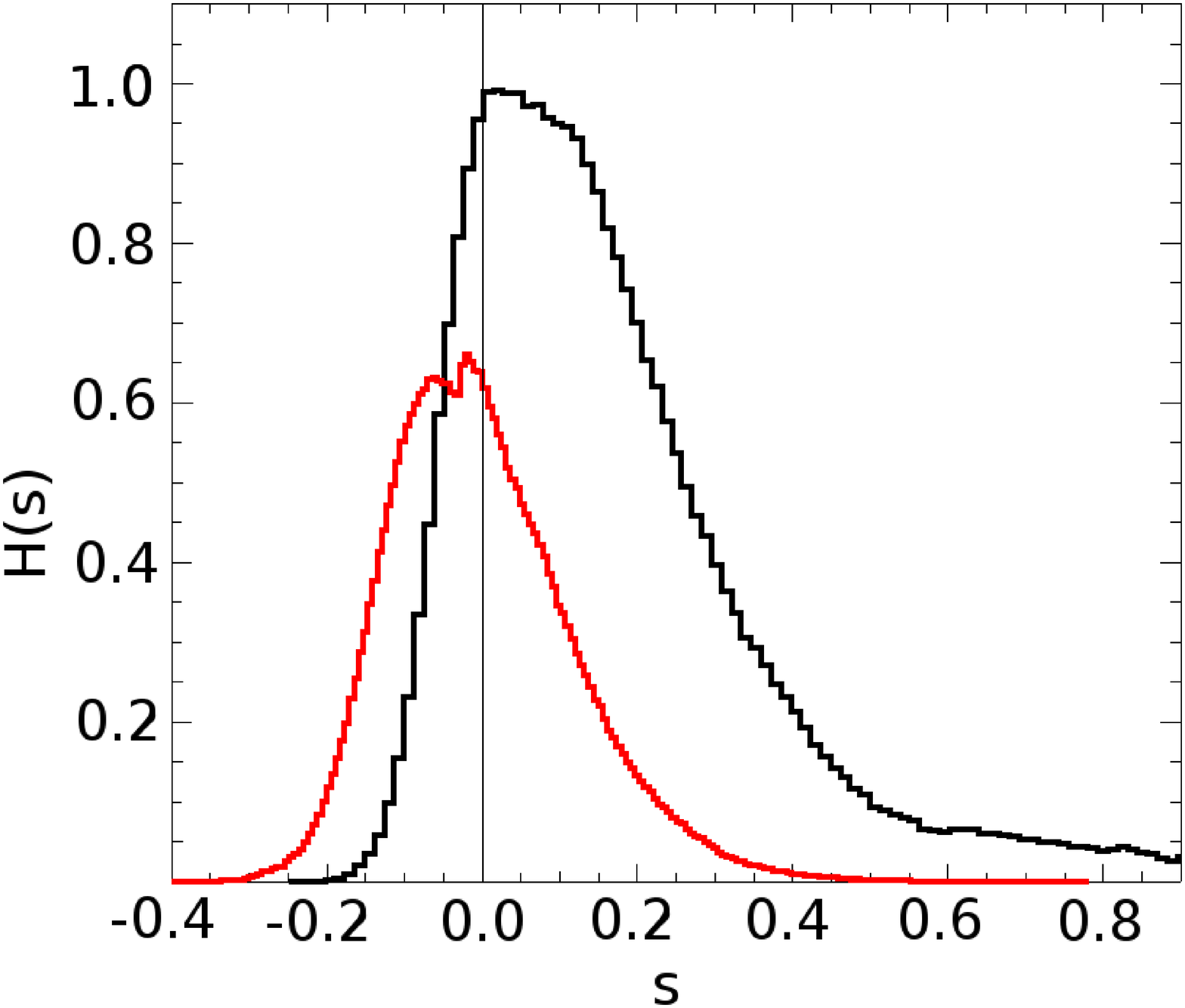}
  \caption{(Color online) Top panel: power spectra for the map from Fig. \ref{fig_Mollweide_map}. 
The black line is the total power spectrum $C^S(l)$, the blue dots are for the
$m=0$ mode only and the red line is for the $|m|\geq 1$ modes. Bottom panel: the normalized distribution function $H(s)$ for 
versus amplitude of fluctuations $s$ for the total signal (the black line) and for the signal with $|m|\geq 1$
(the red line).}
  \label{fig_powerspec_background_map}
\end{figure}
%%%%%%%%%%%%%%%%%%%%%%%%%%%%%%%%%%%%%%%%%%%%%%%%%%%%%%%%%

To further quantify this separation of background from fluctuations, we discuss now in more detail the spherical
 harmonic analysis of $f(\theta,\phi)$.
Using Eq. (\ref{decomposition}) and, for instance, the GLESP transform method,
 we can decompose this distribution into its multipole components (see Appendix A for details). The result of this 
operation is the set of amplitude coefficients $a_{l,m}$. With a suitably high choice of $l_{max}$ all 
morphological features of the initial angular distribution of counts are preserved. 
For the decomposition we have used $ l_{max}=100 $ in accordance with 
the original binning of the fluctuations shown in Fig.~\ref{fig_Mollweide_map}.
The corresponding power spectrum $C(l)$ is shown in the top panel of Fig.~\ref{fig_powerspec_background_map}.

In the case of a perfect reflection symmetry of particle production around mid-rapidity,
 $ \theta - \frac{\pi}{2}   \leftrightarrow - \left( \theta - \frac{\pi}{2}  \right)$, 
the odd components $C^S(l)$ of the total power spectrum vanish.
 Event-by-event fluctuations, 
however, will break this reflection symmetry. As a consequence of such fluctuations,
the top panel of Fig.~\ref{fig_powerspec_background_map} shows 
very small but non-vanishing values $C^S(l) \in \left[10^{-5}; 10^{-4}\right]$
 for essentially all odd integers $l$. In contrast, the even modes of $C^S(l)$ 
are for small even integers $l$ up to a factor 1000 larger since they are dominated by the strong 
dependence of particle production on pseudorapidity. 
Removing the dominant $m=0$ mode from this power spectrum, we see that all even and odd modes 
in $C^S(l)$ have now approximately equal signal strength, 
and that the odd modes do not change in strength. This quantifies the extent to which 
removing the $m=0$ mode removes the trivial background dependence without affecting information on fluctuations. 
 We conclude that
this procedure allows one to represent a 'baseline' event, in which fluctuations
are superimposed on top of a varying background, in a way in which the varying background is efficiently removed and fluctuations dominate the visual representation
(see Fig.~\ref{fig_Mollweide_map}) and the statistical information (see Fig.~\ref{fig_powerspec_background_map}).
 We shall discuss in the following section how this
changes if the event contains, for example, a collective flow component.

We finally mention yet another way of checking that removal of the $m=0$ mode 
isolates information about the nature of fluctuations in the event.
To this end, we recall that $a^*_{l,m}=(-1)^ma_{l,-m}$, and we write the power spectrum of Eq.~(\ref{pow}) in the form
\begin{equation}
 C^S(l)=C(l)+D(l)=\frac{|a_{l,m=0}|^2}{2l+1}+\frac{2}{2l+1}\sum_{m=1}^l|a_{l,m}|^2.
\label{pow1}
 \end{equation}

The first part of Eq.~(\ref{pow1}) , $C(l)$, corresponds to the most symmetric $m=0$ azimuthal  modes, while the second part,
$D(l)$, shows
the power spectrum of fluctuation above the $|m|=0$ threshold. 
It would be worth to note, that both these components have
significantly different statistical properties, which can be characterized in terms of the probability distribution
 functions.
The bottom panel of Fig. \ref{fig_powerspec_background_map} shows the difference in the statistical properties of the
signal when all $m$ modes are included (black) and the case $|m|\geq 1$ (red).
  Normalized to the total number of pixels ( $N_{tot}\simeq 8l^2_{max}$), the probability distribution function, $H(s)$, 
is defined as the number of pixels with corresponding amplitude of fluctuation
within an interval  $[s-\delta s,s+\delta s]$. Here, $\delta s=(s_{max}-s_{min})/N_{bin}$,  $s_{max}$ and $s_{min}$ are
the absolute maxima and minima of the signal in the map with $m \geq 1$ (red), and $N_{bin}=200$ is a number of intervals 
( bins).
 As it is seen from  Fig. \ref{fig_powerspec_background_map}  the probability distribution function for signal with 
removed  the $m=0$ mode,  is quite close to a Gaussian distribution, 
while the one including all $m$ modes (black) is far from a symmetric Gaussian distribution with respect to $s=0$.

To the extent to which the $m=0$-mode dominates the distribution, any distortion of the morphology of
 the map will provide relatively small corrections 
to the total power spectrum $C^S$, but it may result in strong modifications of the power spectrum $D(l)$ 
(see Eq.~(\ref{pow1})), and of the corresponding multipole coefficients $a_{l,m}$.
 Note that any features of the $a_{l,m}$ and of the corresponding power spectrum 
can be detected only if they exceed the level of statistical event-by-event fluctuations. 
This is why the above mentioned 'Gaussianisation' trend of the probability distribution function
for the sub-dominant component power spectrum, $D(l)$, reflects the level of detectability of particular 
features with small amplitudes above the statistical level.

%%%%%%%%%%%%%%%%%%%%%%%%%%%%%%%%%%%%%%%%%%%%%%%%%%%%%%%%%%%%%%%%%
\section{Single event $v_{n}$-flow \label{sec:anisotropies} }
Due to Lorentz contraction the colliding ultra-relativistic nuclei appear pancake-like just before a heavy ion collision in the laboratory frame of reference. For non-central collisions, the overlap between the two nuclei has an approximately almond-shaped cross section. This initial and anisotropic collision geometry results in large pressure gradient differences along the principal axes of the distribution and gives rise to a fluid dynamical flow characterized by fundamental properties of the fluid like the reinteraction time and residual interaction between the constituents. This flow, in turn, leads to an anisotropic distribution of particle momenta and numbers in the transverse plane which may be quantified as a function of the azimuthal angle, $\phi$~\cite{Ollitrault1992, VPS-review2010}. 

Recent experimental results from both RHIC and the LHC suggest that the particle flow is established at the quark-gluon level over a characteristic time scale of about $1-2 fm/c$ and that the flow is quite sensitive to detailed features of the system, such as the viscosity \cite{Teaney:2003kp, Romatschke2007}. Recently,
the presence of higher order flow moments was understood as resulting from fluctuations around the almond shape in the initial collision 
geometry~\cite{Alver-Roland-2010,Sorensen2010,PHENIX-vn-flow,ALICE-vn-flow,Teaney:2010vd,Alver:2010dn,Bhalerao:2011yg}.

In the present study, the angular distribution of particles produced in each heavy ion collision will be viewed as a 'stochastic' map
\begin{eqnarray}
 S(\theta,\phi)=f(\theta,\phi)\left[1+2\sum_n v_{n}cos[n(\phi-\Psi_n)]\right]\, .
\label{eq6}
\end{eqnarray}
This expression maps the  random distribution  of particles without flow, $f(\theta,\phi) \equiv {d^2N}/{d\phi d\eta}|_{v_{n} = 0}$ onto a distribution  
$S(\theta,\phi)= d^2N/ d\phi d\eta$ that shows azimuthal dependencies with harmonic flow amplitudes $v_{n}$ in azimuthal orientations $\Psi_n$. 
The model distributions considered in the following were obtained in line with this expression 
by modulating, a posteriori, an azimuthal uniform distribution $f(\theta,\phi)$ simulated
in HIJING or a simple random generator with prescribed flow amplitudes $v_{n}$ with angular orientations $\Psi_n$~\footnote{
 In general, the event plane angles $\Psi_n$ are correlated to the azimuthal orientation $\Psi_R$ of the 
reaction plane, that is defined as the plane spanned by the impact parameter vector and the beam direction. However, 
while $\Psi_R$ and the $\Psi_n$'s are correlated, they are in general not identical and different harmonic modes can be correlated differently.
Therefore, there will be in general an event plane angle $\Psi_n$ associated with each harmonic $n$, that may be determined in a more comprehensive analysis in terms of 
higher order harmonics, and different $\Psi_n$'s will not coincide in general.}.
The case $n=2$ corresponds to so called {\it elliptic flow}, to which we restrict ourselves in the present study. For simplicity, the model used in this section assigns
the same flow value to all regions of $\eta$ and transverse momentum $p_t$, and to all particle species  (e.g. mesons and baryons). This could be
extended easily to account for dependencies in these parameters, or to treat the flow coefficients as random functions with statistical properties that
differ from those of  $f(\theta,\phi)$. The arguments in the present study will not require such a more detailed modeling. 

A typical experimental task is to extract from a given particle distribution $d^2N/ d\phi d\eta$ the flow coefficients $v_{n}$,
 defined as~\cite{Voloshin-Zhang-1996}
% {\bf Urs has removed a square on the rhs of this equation}
%
\begin{eqnarray}
 \{v_{n}\}=\langle\langle e^{in(\phi-\Psi_n)}\rangle\rangle\, .
\label{eq5}
\end{eqnarray}
Here $\langle\langle..\rangle\rangle$ denotes the average over all particles for each heavy ion collision and over the entire statistical ensemble of events. 
The following discussion will not involve this later step. Rather, we shall discuss how different flow coefficients $v_{n}$ and event planes $\Psi_n$ can be 
identified in a CMB-like harmonic analysis of \textit{single heavy ion events}. To this end, we decompose Eq.(\ref{eq6}) in spherical harmonics 
\begin{eqnarray}
 b_{l,m}\simeq a_{l,m}+\sum_nv_n\left(c_{l,m+n}e^{-in\Psi_n}+c_{l,m-n}e^{in\Psi_n}\right)\, .
\label{eq8}
\end{eqnarray}
Here, $b_{l,m}$ are the coefficients of the spherical harmonic decomposition for $S(\theta,\phi)$, and $c_{l,m\mp   n}=a_{l,m}g(l,m,\mp n)$, where
$g(l,m,\mp n)=2\pi N_{l,m\mp n}N_{l,m}\int^1_{-1}dx P^{m\mp n}_l(x)P^m_l(x)$.
This type of equation is also encountered  in the CMB data analysis~\cite{CMB_modulation_analysis} 
for cases where the statistical isotropy of the signal is broken by regular modulations. We now discuss this equation in more detail.

%%%%%%%%%%%%%%%%%%%%%%%%%%%%%%%%%%%%%%%%%%%%%%%%%%%%%%%%%%%%
\begin{figure}[!h]
 \begin{center}
  \centerline{\includegraphics[scale=0.35]{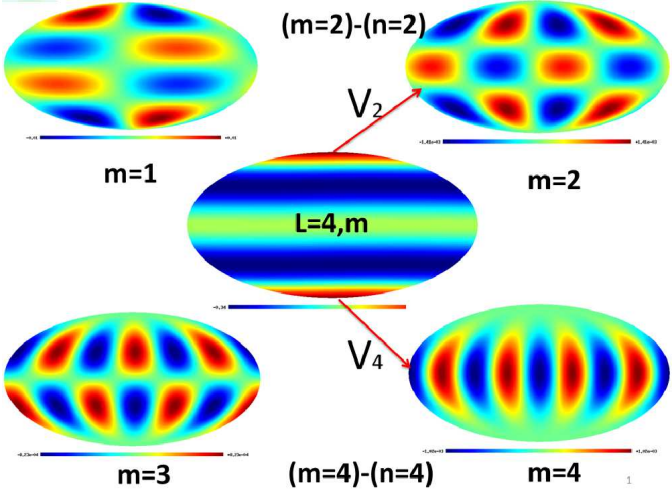}}
  \caption{(Color online) Schematic representation of the contribution of $v_2$ (elliptic) and $v_4$ flow to the various $m$ components of the $b_{l=4,m}$-coefficient.
The $v_2$-modulation will change the amplitude and the phase of the $b_{4,2}$ coefficient % through the $m=2\leftrightarrow n=2$ coupling, 
without contributing to the other components. The $v_4$ modulation will only change the $b_{4,4}$ component.}
% through the $m=4\leftrightarrow n=4$ coupling.}
  \label{fig_v2_v4_flow_schematic}
 \end{center}
\end{figure}
%%%%%%%%%%%%%%%%%%%%%%%%%%%%%%%%%%%%%%%%%%%%%%%%%%%%%%%%%%%%%%%

\subsection{$v_n$-modulation with $n=even$ \label{subsec:even-n} }

The iterative solution of Eq. (\ref{eq8}) has an especially simple form for even-$n$, $n=2k, k=1,2..$. In this case, the
dominant component of the $a_{l,m}$-coefficients is associated with $m=0$ modes, and $l=even$.
As we have pointed out in Section 3, the strongest component of the $f(\theta,\phi)$ signal is the $m=0$ component (blue dots in Fig. \ref{fig_powerspec_background_map}). This implies that, in 
Eq. (\ref{eq8}), the last term in the brackets $a_{l,m-n}$ is maximal, if $m$.

For $m=0$, Eq. (\ref{eq8}) reads 
\begin{eqnarray}
 b_{l,0}&=&a_{l,0}+\sum_nv_n\left(c_{l,n}e^{-in\Psi_n}+c_{l,-n}e^{in\Psi_n}\right)\nonumber\\
&=&a_{l,0}+2\sum_nv_n|c_{l,n}|\cos(n(\phi_{l,n}-\Psi_n))\, ,
\label{eq9}
\end{eqnarray}
where we have used the relation $a_{l,-m}=(-1)^m a^*_{l,m}$.
For the scenario studied here, we can neglect in Eq.~(\ref{eq9}) the second term ($\propto v_n|c_{l,n}|$) relative to  the first one, due to the 
inequality $|a_{l,0}|\gg |c_{l,n}|$ which follows from the dominance of the $m=0$ modes. We consequently obtain the following approximate solution: $b_{l,0}\simeq a_{l,0}$. 

The coefficients $b_{l,m}$ vanish for $l<|m|$. As a consequence, in Eq. (\ref{eq8}), for instance, the coefficient $b_{l,m}$ with $m=2$ and $l=2 $ is connected to the
$v_2$ component only, and does not receive contribution from higher flow momentum with $n>2$. For $l=4$ there exists only two terms in $b_{4,m}$ which 
are in resonance with the $b_{4,0}$ mode: $b_{4,2}$ through the $v_2$ term in Eq. (\ref{eq8}), and $b_{4,4}$ through $v_4$. We illustrate
these relations further in Fig. \ref{fig_v2_v4_flow_schematic}.
We note that the elliptic flow will also provide a $b_{4,3}\leftrightarrow b_{4,1}$
coupling, but this effect is significantly smaller ($\sim v_2$), in comparison with the discussed $b_{4,2}$ and $b_{4,4}$ components.
These features of Eq. (\ref{eq8}) motivate to use 
\begin{eqnarray}
 b_{n,n}=a_{n,n}+ v_nb_{n,0}g(n)e^{in\Psi_n},
\label{eq10}
\end{eqnarray}
as a basis for separating the contributions from different $v_n$, 
where $g(n)=2\pi N_{n,0}N_{n,n}\int^1_{-1}dx P^{0}_n(x)P^n_n(x)$ (see Appendix A).

For a sufficiently large flow signal, such that $|a_{n,n}|\ll v_n|b_{n,0}|g(n)$, 
the iterative solutions of Eq.~(\ref{eq10}) for the amplitude of the flow $v_n$ and the event plane $\Psi_n$ for a single event are then given by

\begin{eqnarray}
 v_n\simeq \frac{|b_{n,n}|}{g(n)|b_{n,0}|},\hspace{0.5cm}n\Psi_n=\phi_{n,n}.
\label{eq11}
\end{eqnarray}

\begin{figure}[!t]
 \begin{center}
  \centerline{\includegraphics[scale=0.7]{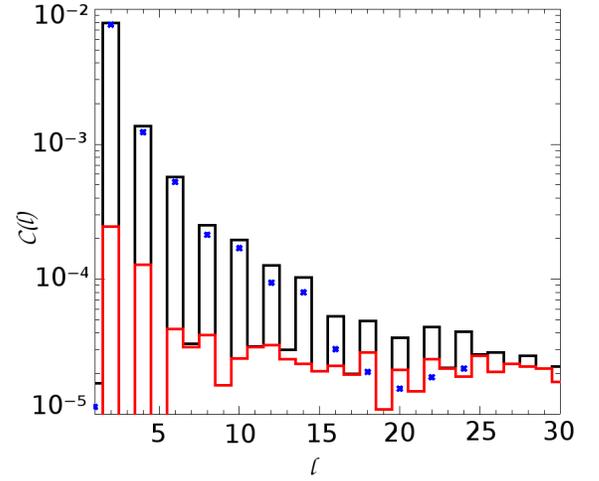}}
  \caption{(Color online) The power spectrum for an event with elliptic flow $v_2=0.07$. The total power $C^S(l)$ is shown in black,
 the blue stars correspond to $C^S(l)-D(l)=C(l)$, and the $D(l)$ power spectrum is shown in red (see Eq.(3)). }
  \label{fig_powerspec_v2}
 \end{center}
\end{figure}

\subsection{$v_n$-modulation with $n=odd$ \label{subsec:odd-n} }

The method presented above clearly illustrates how spherical harmonics with $l=even$ give access to amplitudes of the $v_n$-flow coefficients with $n=2,4,6.. $. 
We now focus on the odd harmonics $v_n$, $n=1,3,5, ...$ that are known to provide information about the fluid dynamic response to fluctuations present in the
initial stage of heavy ion collisions. 
To this end, we study  the $b_{l,m}$-basis of Eq.~(\ref{eq8}) for the case $l=even$ and $n=odd$.
We start from the most powerful $l=2$ and $l=4$ components of $b_{l,m}$ under consideration (see Fig.~\ref{fig_powerspec_background_map}).
For the quadrupole component, $l=2$, in the presence of even and odd flow harmonics, the ''resonant'' component $b_{2,m}$ for $v_1$ is $b_{2,1}$:

\begin{eqnarray}
 b_{2,1}=%a_{2,1}+v_1(c_{2,0}e^{i\Psi_1}+a_{2,2}e^{-i\Psi_1})+v_2a_{2,-1}e^{2i\Psi_2}\nonumber\\
&\simeq& v_1c_{2,1}e^{i\Psi_1} %+a_{2,1}(1-v_2e^{2i\Psi_2}),
\label{odd1}
\end{eqnarray}
where  the phase of the event plane $\Psi_1$  corresponds to $v_1$-flow.
In Eq. (\ref{odd1}) the major contribution to $b_{2,1}$ is given by the first term, proportional to the most powerful component $a_{2,0}\simeq b_{2,0}$, and

\begin{eqnarray}
 b_{2,1}\simeq v_1b_{2,0}g_1e^{i\Psi_1},\hspace{0.3cm}v_1=\frac{|b_{2,1}|}{g_1|b_{2,0}|},\hspace{0.3cm}\Psi_1=\phi_{2,1}\, ,
\label{odd2}
\end{eqnarray}

\noindent where $\phi_{2,1}$ is the phase of the $b_{2,1}$-component, and $g_1=4\pi\int_0^1dxP_2^0(x)P_2^1(x)$.
It is obvious that for any odd $n$ the corresponding solution for the amplitude and the phase is given by

\begin{eqnarray}
 v_n=\frac{|b_{n+1,n}|}{g(n+1)|b_{n+1,0}|},\hspace{0.3cm} n\Psi_n=\phi_{n+1,n}\, .
\label{odd3}
\end{eqnarray}
where $g(n+1)=4\pi\int_0^1dxP_{n+1}^0(x)P_{n+1}^{n}(x)$. Similar to $n=even$ case, the phase of the event plane for $n=odd$ can be reconstructed also from $\phi_{n,n}$ phase as follows: $n\Psi_n=\phi_{n,n}$.\\
 
Note that all methods of reconstructing the orientation of the reaction plane(s) in heavy ion collisions distinguish between the 'true'
 reaction plane $\Psi_n$ and the orientation of the event plane reconstructed from experimental data $\Psi_n^{exp}$. 
Uncertainties in extracting the true parameter from a statistically finite amount of data are characterized by quoting the finite resolution.
That means that, in respect to the statistical ensemble of realizations, $v_n$ and $\Psi_n$ should be treated as 
random variables with probability density functions $P_{v_n}$ and $P_{\Psi_n}$. Performing the same estimation as
 in Eq. (\ref{eq11}) and Eq. (\ref{odd3}) for each event we can determine the number of realizations with $v_n$ and $\Psi_n$
 within a given interval of uncertainty, and in fact, we can obtain the probability distribution functions $P_{v_n}$ and $P_{\Psi_n}$ 
for a given ensemble of realizations.
 Then, one can define the first moments of the corresponding variables $\langle v_n\rangle$ and $\langle \Psi_n \rangle$, for a statistically finite amount of data. 
We will illustrate this approach in the next section.

\begin{figure}[!h]
 \begin{center}
  \centerline{\includegraphics[scale=0.15]{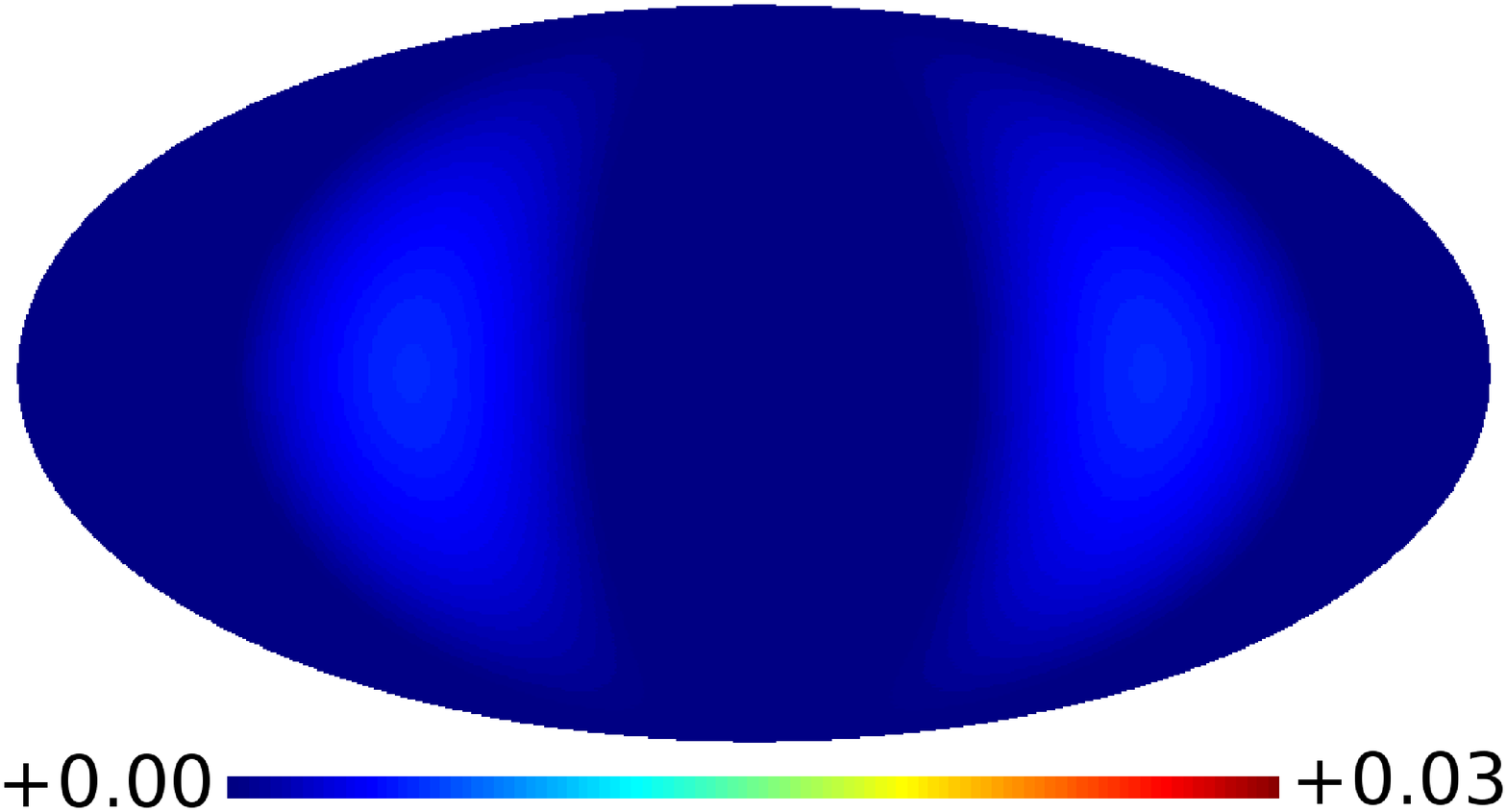}%}
		\includegraphics[scale=0.15]{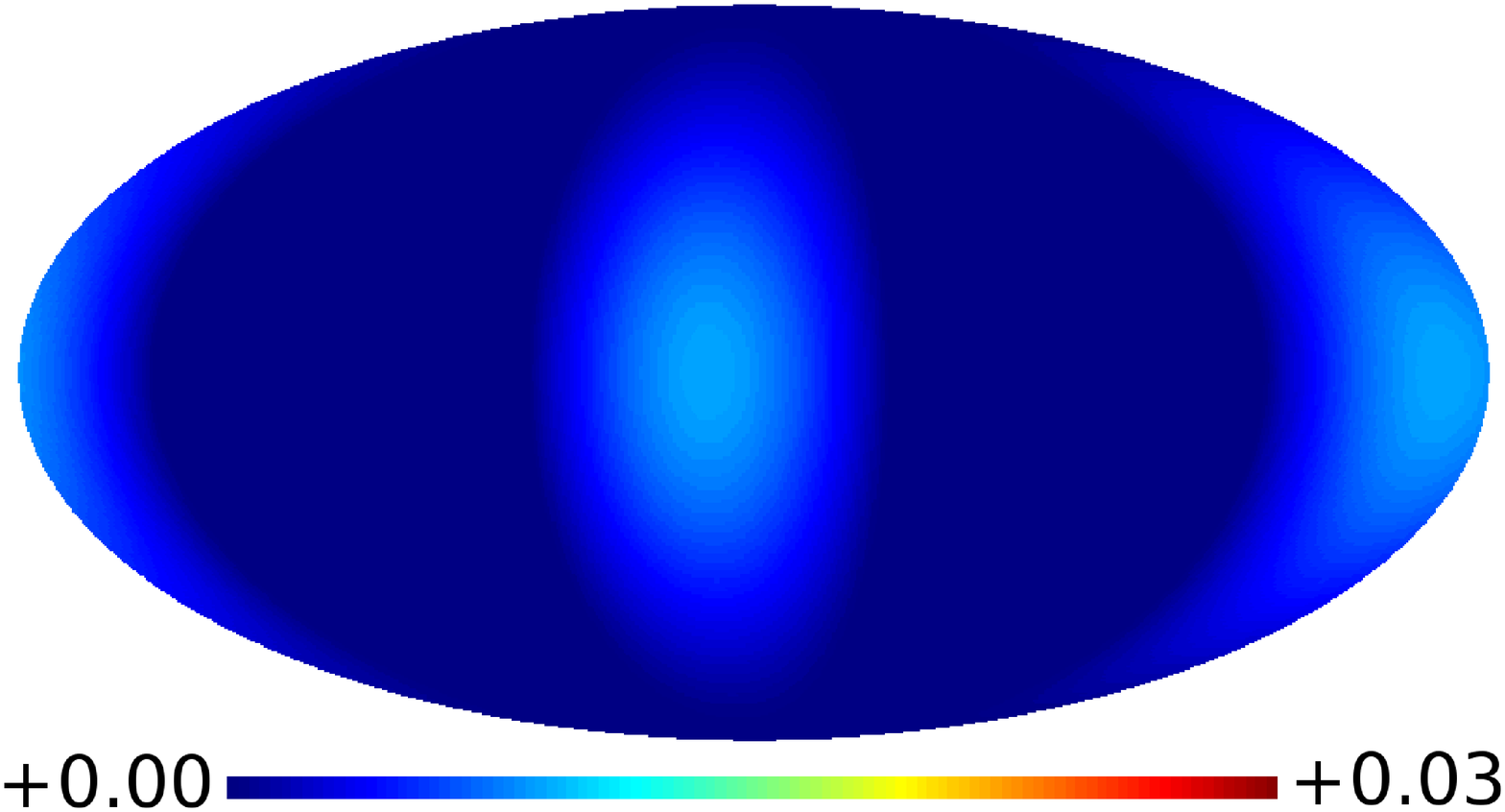}}
  \centerline{\includegraphics[scale=0.15]{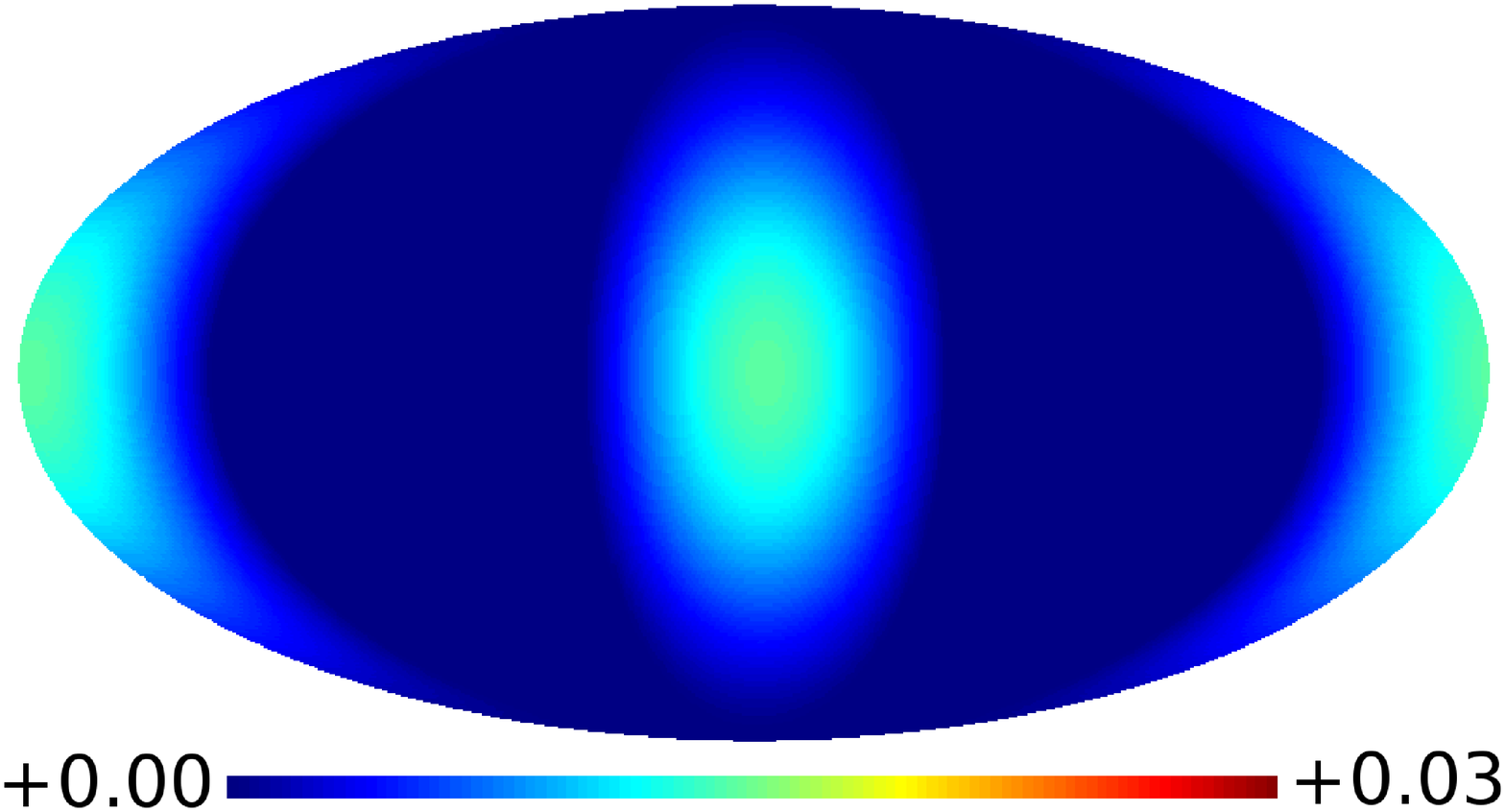}%}
		\includegraphics[scale=0.15]{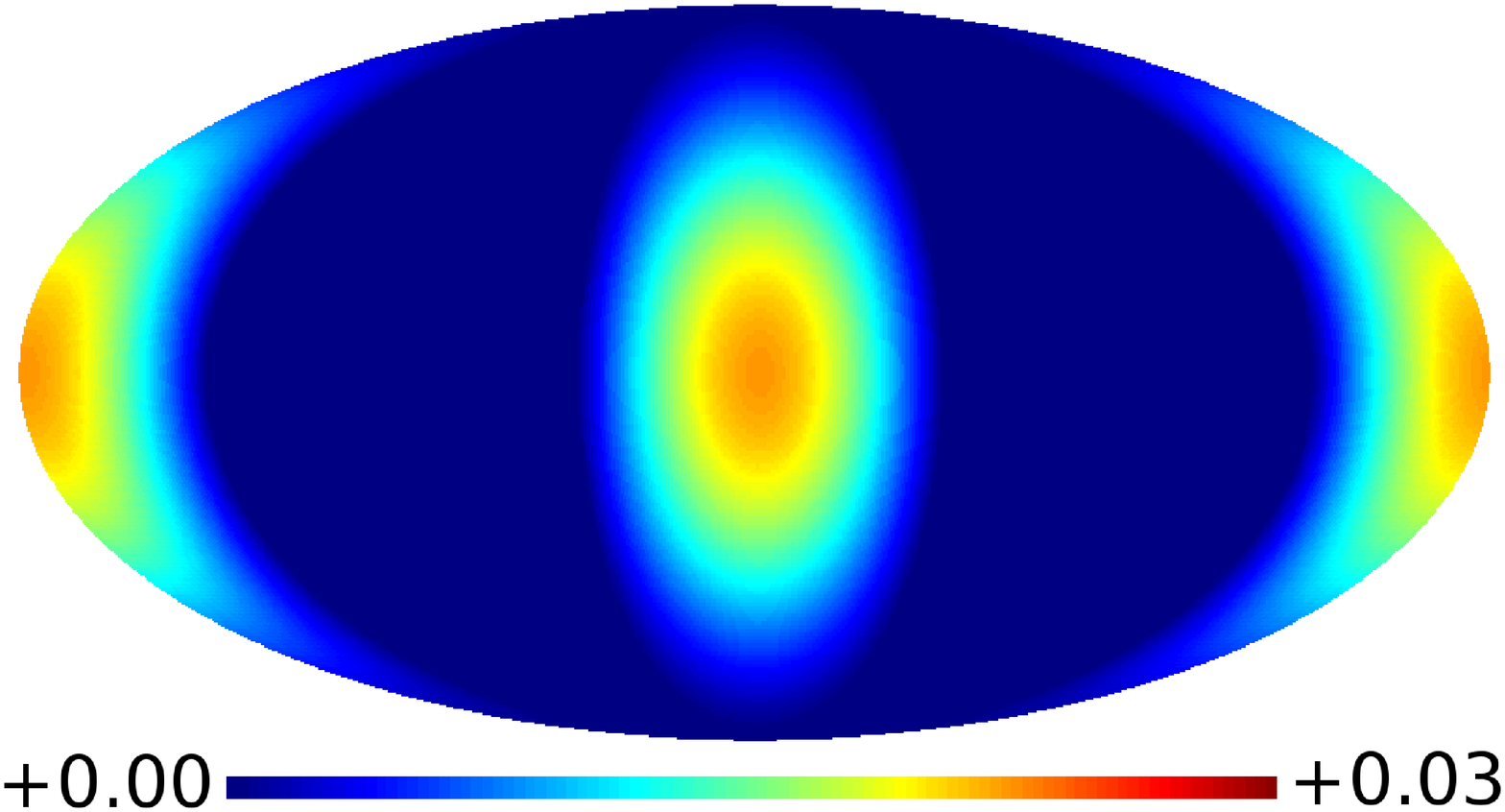}}
  \centerline{\includegraphics[scale=0.15]{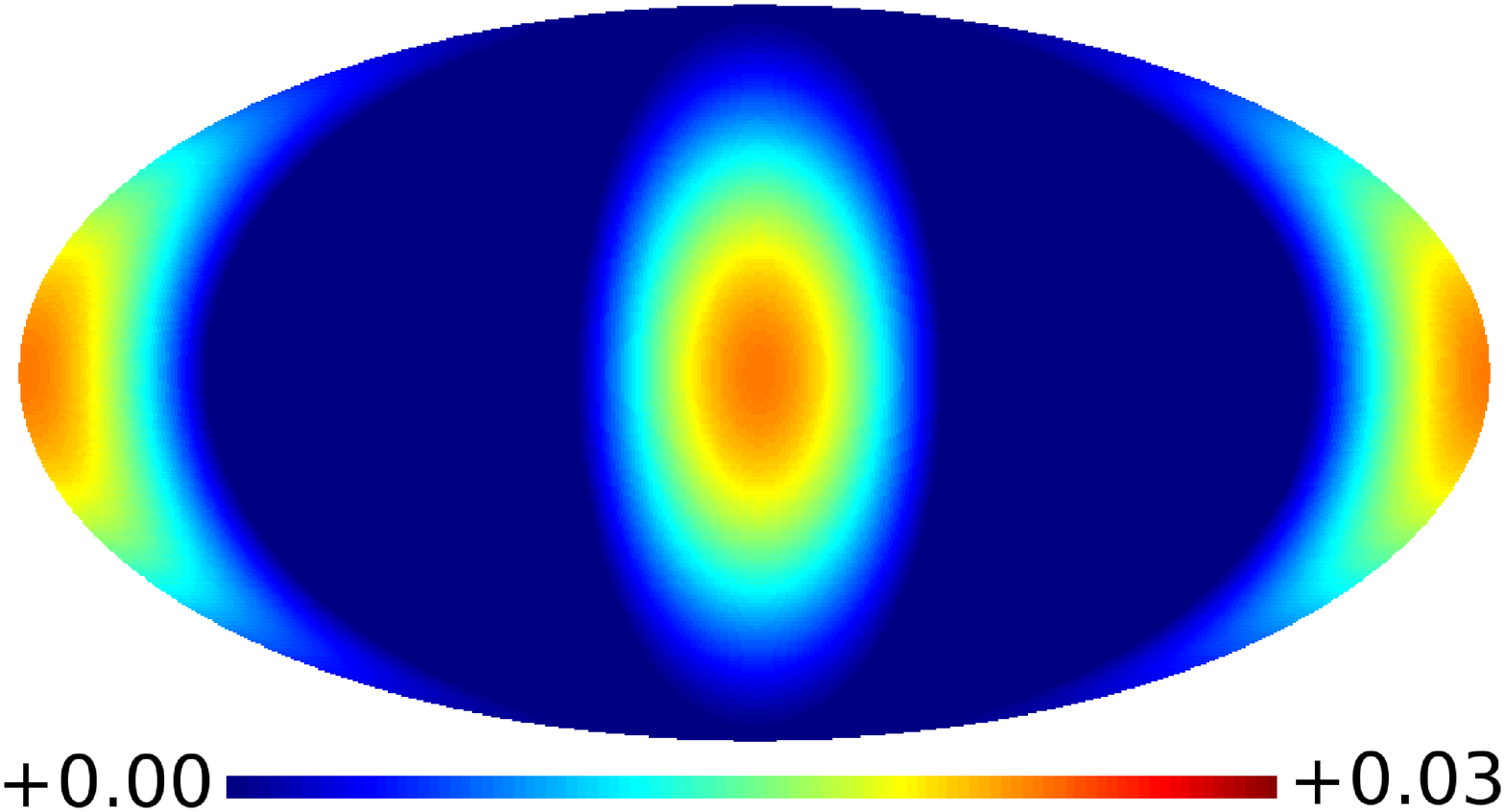}%}
		\includegraphics[scale=0.15]{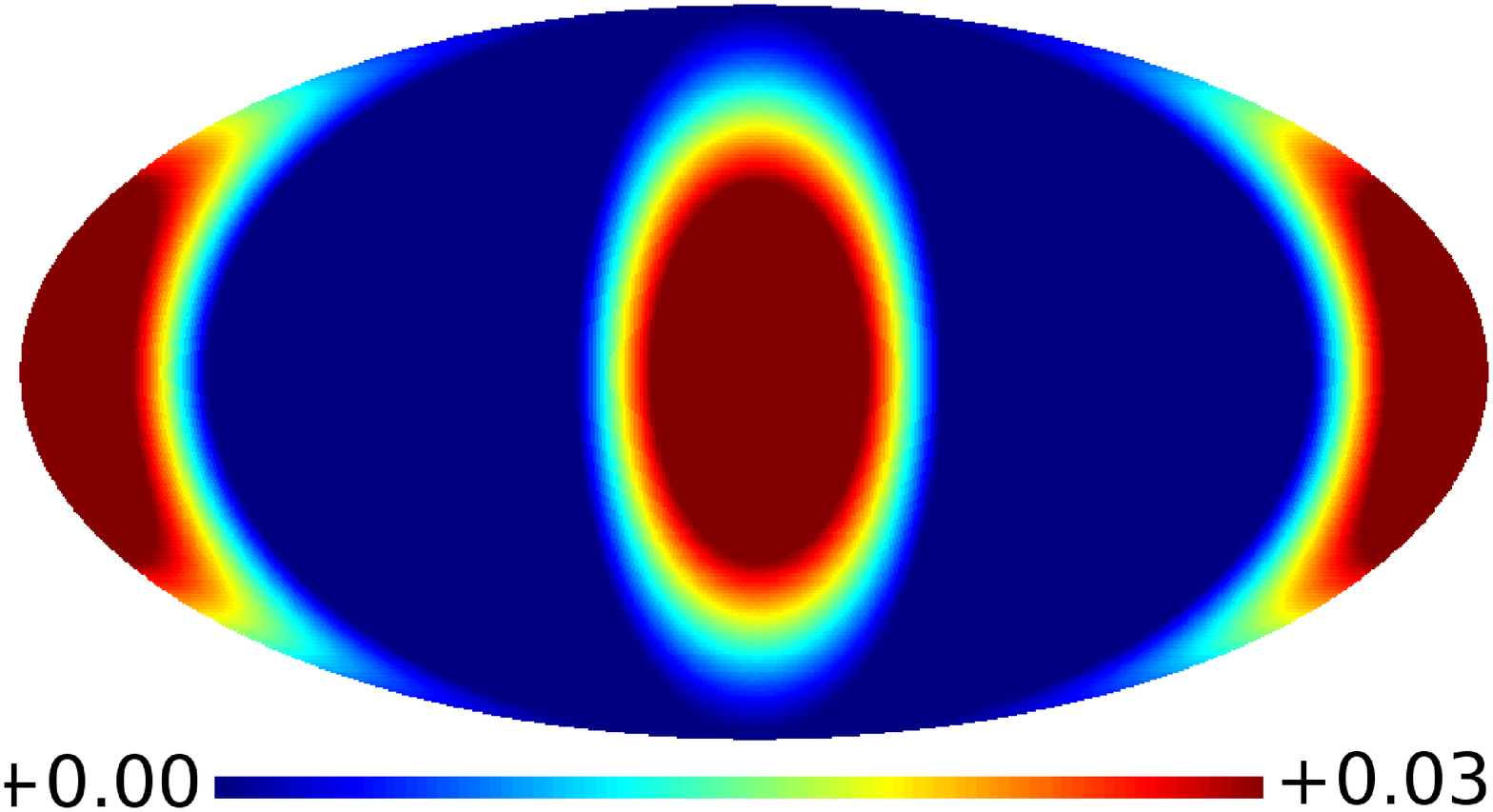}}
  \centerline{\includegraphics[scale=0.15]{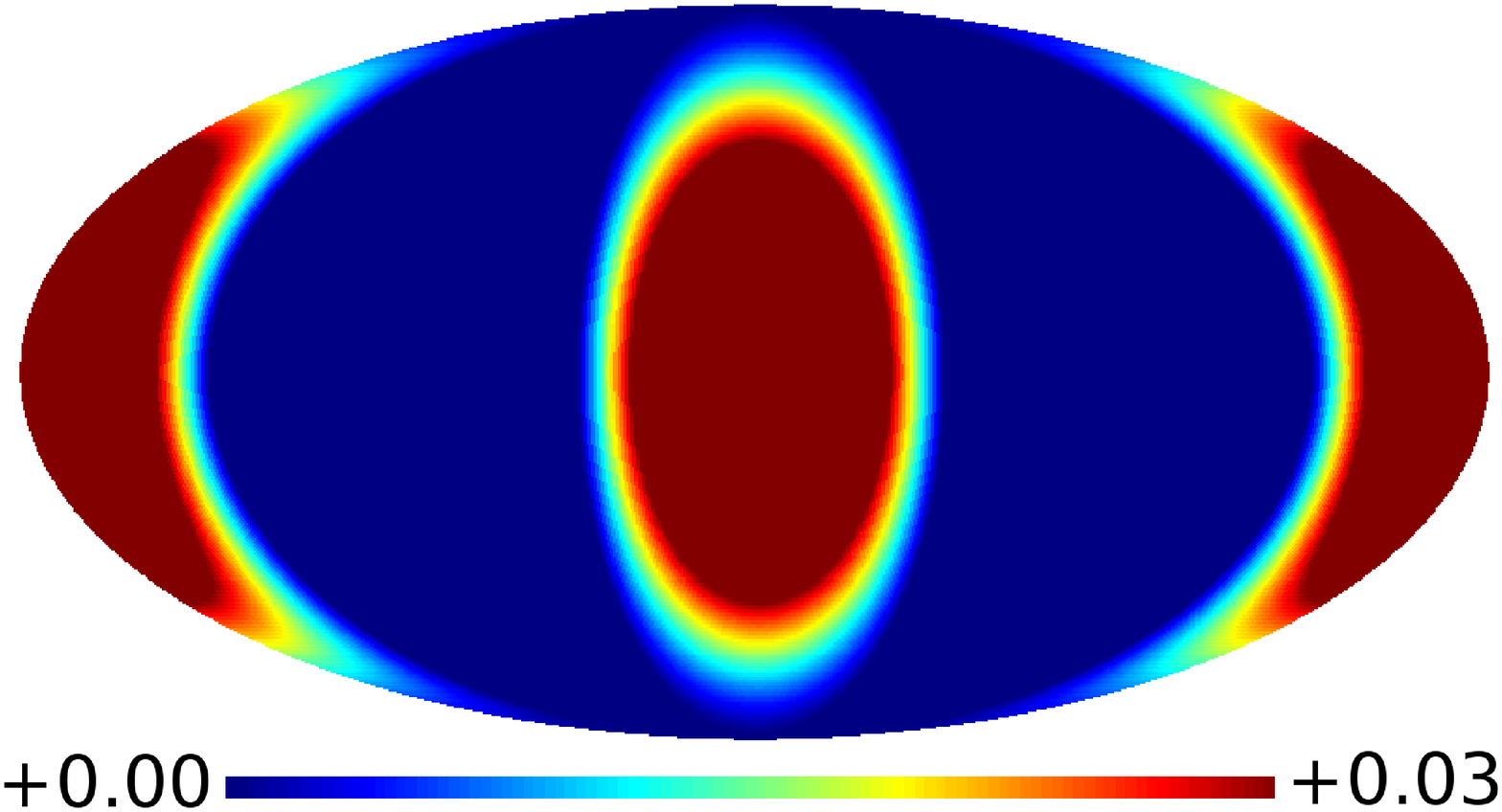}%}
		\includegraphics[scale=0.15]{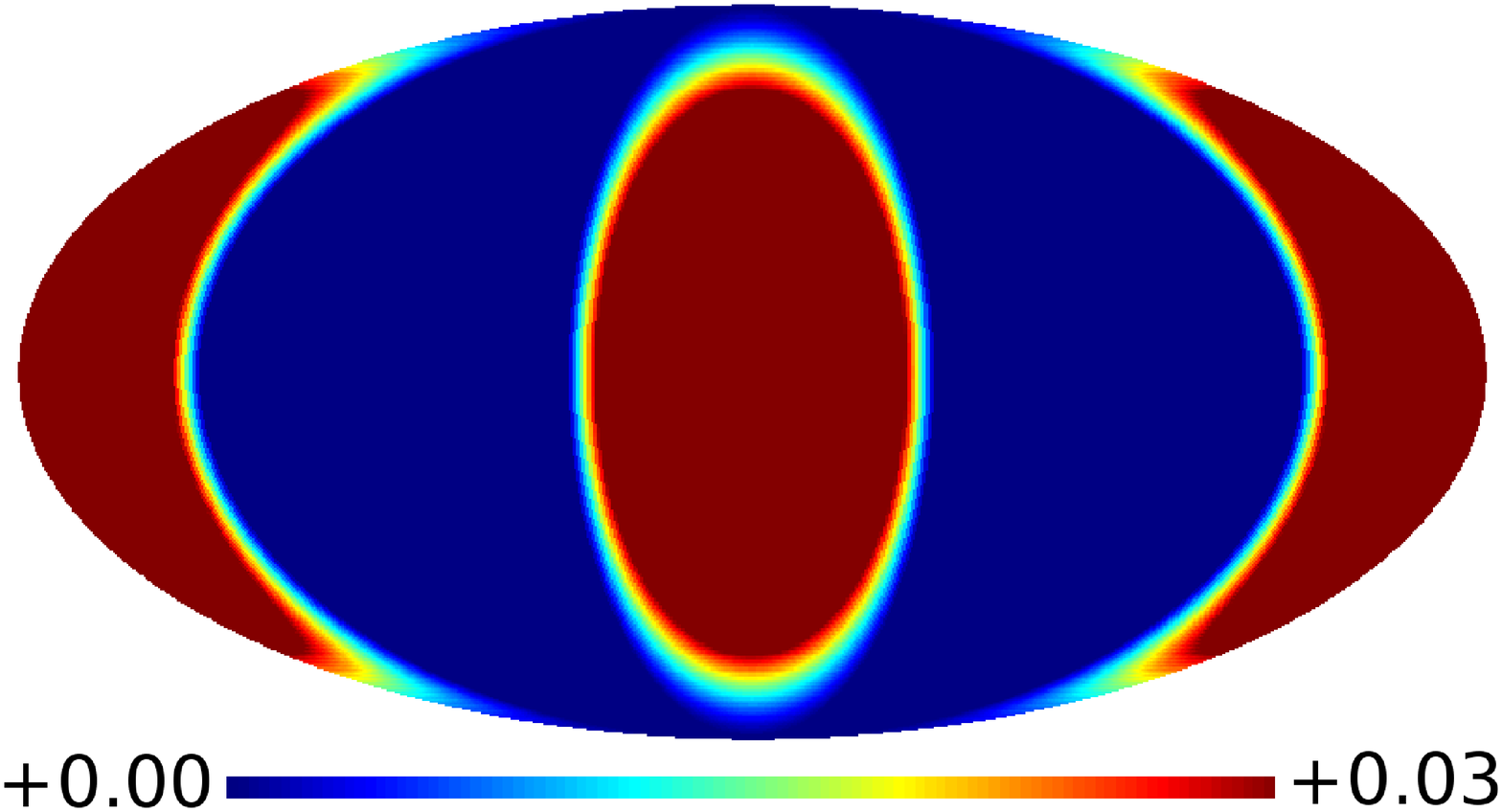}}
  \caption{(Color online) From the top left down to the bottom right: Maps of the $|b_{22}|$ amplitude for input flow values of $v_2=0,0.01,0.05,0.075,0.1,0.2,0.3,0.5$, respectively. For simplicity $\Psi^2_R=0$ in all maps. A different reaction plane angle would manifest itself as a phase shift in the 'East-West' direction.}
  \label{fig_maps_b22_amplitudes}
 \end{center}
\end{figure}

%%%%%%%%%%%%%%%%%%%%%%%%%%%%%%%%%%%%%%%%%%%%%%%%%%%%%%%%%%%%%%%%
\section{Elliptic flow: the case $n=2$ \label{sec:ellipticflow} }

Here, we further illustrate how the symmetries of the spherical harmonic representation of the particle distribution 
can be used to extract the elliptic flow signal following Eq.~(\ref{eq11}). As discussed in Section 3 and shown in Fig. \ref{fig_dNdPhi}, this signal
is a modulation of the random background particle distribution by a factor 
$ \propto v_n\cos(2\phi-\Psi)$. In Fig. \ref{fig_powerspec_v2} we show the corresponding power spectrum using a flow of $v_2=0.07$. 
The power in the components $D(l)$, $l=2,4,6,8$ is seen to be significantly higher than the asymptotically flat and small $D(l)\simeq const$ values, 
obtained from a random realization without flow (see Fig. \ref{fig_powerspec_background_map}). It is in this way that the CMB-like harmonic analysis
identifies - on the basis of a single event and without recourse to an event sample - the presence of a signal on top of random localized fluctuations. 
The enhancement of the even low-l multipoles originates from flow contributions to $b_{2,2}$, $b_{4,2}$, $b_{6,2}$. Expressing these multipoles
via  Eq.~(\ref{eq8}), we find 
\begin{eqnarray}
 D(l)&=&\frac{1}{2l+1}\sum_{m=1}|a_{l,m}+v_2b_{l,0}g(n)\delta_{m,n}e^{in\Psi_n}|^2\nonumber\\
&\simeq& v^2_2C(l)g^2(n)+\overline D(l)\, ,
\label{eq12}
\end{eqnarray}
where $(2l+1)\overline D(l)=\sum_{m=1}|a_{l,m}|^2$. 

%%%%%%%%%%%%%%%%%%%%%%%%%%%%%%%%%%%%%%%%%%%%%%%%%%%%%%%%%%%%%%%%
\begin{figure}[!t]
 \begin{center}
  \centerline{\includegraphics[scale=0.53]{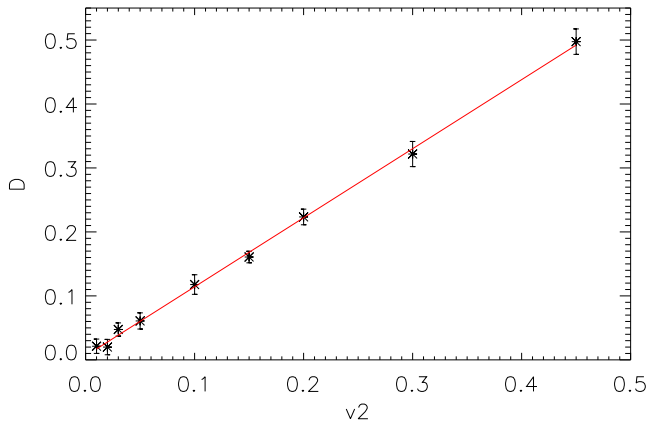}}
  \caption{(Color online) Estimator $D=|b_{2,2}|/g(2)|b_{2,0}|$ for the elliptic flow amplitude obtained for single events plotted versus the $v_2$ value used in the  simulations. The error bars correspond to $68\%$ confidence level.  }
  \label{fig_b22_v2}
 \end{center}
\end{figure}

\begin{figure}[!t]
 \begin{center}
 \centerline{\includegraphics[scale=0.33]{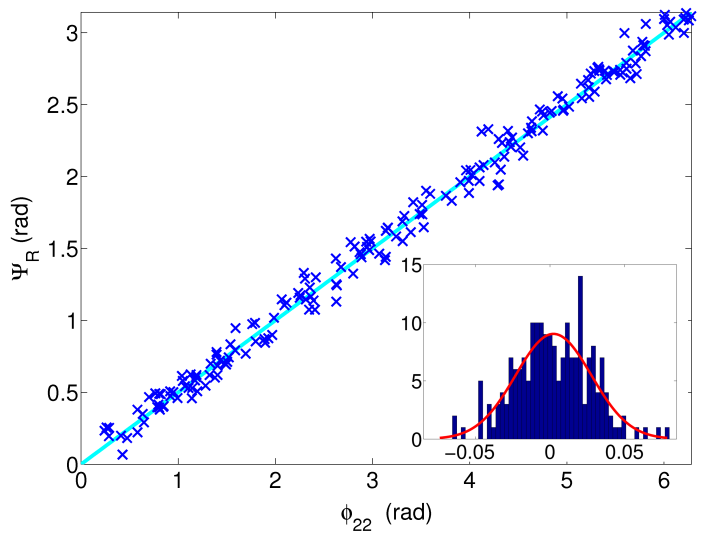}}
  \caption{(Color online) The reaction plane angle $\Psi_R$ (rad) for each of 202 HIJING events with flow = 0.07 versus the phase $\phi_{2,2}$ (rad) obtained from the analysis described in Eq. (\ref{eq11}). The insert shows the distribution of the difference $\Psi_R - 0.5 \phi_{2,2}$ (rad).}
  \label{fig_PsiN_vs_PhiNN}
 \end{center}
\end{figure}

We discuss now in further detail the relation between the flow signal and the multipole components determined in a CMB-like harmonic analysis. 
In a Mollweide projection, one can visualize easily how the $|b_{2,2}|$ component changes with a flow signal 
varying between  $v_2=0 $ to $v_2=0.5 $, see  Fig. \ref{fig_maps_b22_amplitudes}. In Fig. \ref{fig_b22_v2}, we demonstrate this dependence beyond
the level of single events by averaging over samples of $10^5$ events with a specific flow amplitude $v_2$. We see that there is a linear
relation between $v_2$ and the ratio of the dominant quadrupole components  $|b_{2,2}|$ and $|b_{2,0}|$. The small error bars shown in 
Fig. \ref{fig_b22_v2} correspond to a $1\sigma$ standard deviation. They illustrate the degree to which in the present model study fluctuations 
at the single event are small compared to the event-averaged mean. An analogous conclusion can be drawn from 
Fig. \ref{fig_PsiN_vs_PhiNN} about the correlation between the true event plane ($\Psi_R$), which is known for each simulated event,
and the phase $\phi_{2,2}$ determined in the analysis (see Eq. (\ref{eq11})). The insert shows the distribution of the difference 
$\Psi_R - 0.5 \phi_{2,2}$ which is a measure of the event plane resolution obtained in this method. The distribution is consistant with a Gaussian 
with  mean value $\mu \approx 0$ and  standard deviation $\sigma =0.0254 rad.$ 
The tight correlation between both orientations is due to the fact that particles at all 
rapidities show on average the same azimuthal correlations and this contributes to constraining statistical fluctuations in $\phi_{2,2}$. 

We finally indicate how a CMB-like harmonic analysis can be used to assign to an event a probability that the measured harmonic 
components result as fluctuations from a random background, rather than in response to a collective flow signal. To assess this
question,
 we have generated 
the distribution of 
$|a_{2,2}|$ values obtained by decomposing $10^5$ events without 
flow into spherical harmonics. The probability $P$ of assigning a given event as a random background for, 
say $v_2=0.01$, can be defined as the number of realizations without flow above the 
limit of $|b_{2,2}|$ corresponding to $v_2=0.01$. 
We have found 9230 
out of $10^5$ random realizations, and the  probability of correctly concluding an elliptic flow value of $v_2=0.01$ to an event with this $|b_{2,2}|$ analysis is 
therefore $1-P\simeq 90\%$. For a five times larger signal $v_2=0.05$, the corresponding level of detectability is better than $99,999\%$. 
 We parallel this analysis for the harmonic component $|b_{2,1}|$ and have confirmed
 the fact that the limits for $v_2=0.01$ and $v_2=0.05$  both lie around the average of the background distribution.

%%%%%%%%%%%%%%%%%%%%%%%%%%%%%%%%%%%%%%%%%%%%%%%%%%%%%%%%%%%%%%%%

The exploratory analysis presented here involves information about the simulated events that is not directly accessible experimentally. 
For instance, experimental data do not allow to check directly the correlation between $|b_{2,2}|$ and the 'true flow signal $v_2$', or between $\phi_{2,2}$ and the 'true 
reaction plane $\Psi_R$, as done in  Fig.~\ref{fig_b22_v2} and Fig.~\ref{fig_PsiN_vs_PhiNN}.  In this sense, our discussion serves mainly the purpose to illustrate
that a CMB-like harmonic analysis can be applied to individual heavy ion events and that it has the sensitivity of associating flow values and 
event planes to single events in a meaningful way. We further note that while Figs.~\ref{fig_b22_v2} and ~\ref{fig_PsiN_vs_PhiNN} are not directly
measurable, closely related quantities are. For instance, one could plot the ratio of the quadrupole components $|b_{2,2}|$ and $|b_{2,0}|$
as a function of event centrality to characterize the nature of event-by-event fluctuations in an experimentally measured event sample. 
We also note that we have tested the sensitivity of the method to a coarser binning of the primary particle multiplicity information, as it could result
from a finite binning of the detectors. We find no significant difference in the results when using 100 bins in $\eta$ and 40 bins in $\phi$. Also, methods 
have been developed for the CMB analysis to handle partial detector acceptance and efficiency etc. A more detailed discussion of these aspects will
be left to further work.

%%%%%%%%%%%%%%%%%%%%%%%%%%%%%%%%%%%%%%%%%%%%%%%%
\section{Summary and Conclusions \label{Summary} }

The present work is a bold exploratory step aimed at testing the use of CMB-like analysis tools 
for the study of high-multiplicity heavy ion collisions. It parallels other recent efforts, see e.g.~\cite{Mishra,Mocsy}.
 We started from representing simplified simulations 
of heavy-ion collisions in a Mollweide projection commonly used in the CMB analysis. We then asked
how pertinent features of heavy ion collisions, including collective flow and small-scale fluctuations
manifest themselves visually in such a representation, and how they are characterized by standard CMB tools.

A CMB-like analysis of fluctuations is based on an expansion in terms of spherical harmonics and thus returns
the two-parameter set of coefficients $b_{l,m}$. For the characterization of collective flow phenomena, that 
show up as purely azimuthal dependencies, the azimuthal harmonics with flow amplitudes $v_{n}$
provide a complete set of functions. Therefore, the larger  set of coefficients $b_{l,m}$ must arguably contain redundancies
for the characterization of collective flow in heavy ion collisions. 
However, event-wise fluctuations in heavy ion collisions are not confined to the azimuthal $\phi$-direction and may be expected
to show locally no preferred orientation in the $\eta$-$\phi$-plane. Our study indicates that the larger set $b_{l,m}$, while 
showing redundancies for the characterization of $\phi$-dependent phenomena, may offer novel opportunities towards characterizing 
small-scale fluctuations in heavy ion collisions and separating them from flow effects. For instance, we observed that due to
the $\eta \to - \eta$ reflection symmetry of event-averaged identical nucleus-nucleus collisions, expansion coefficients with even 
integers $l$ are inevitably dominated by event-wise fluctuations. The set of these coefficients may then be used as a baseline
on top of which collective or global event properties can be established. We have discussed all possible generalizations of the method in Appendix A,
focusing on $\theta$-dependency of the amplitude of the  flow .This baseline allowed, for instance, to check in 
Fig.~\ref{fig_powerspec_background_map} 
to what extent removal of the $m=0$ mode subtracts efficiently the effects of a rapidity-dependent multiplicity distribution 
from a fluctuation analysis. And we have seen in the discussion of Fig.~\ref{fig_powerspec_v2} how the larger set of coefficients $b_{l,m}$ can help
visually and statistically to identify flow on top of a fluctuating background. 

For the detection of flow-like azimuthal modulations, the method used here is limited by the level of statistical noise $v_{n}\ge N^{-\frac{1}{2}}$, 
where $N$ is the multiplicity. Thus, for $v_{n}\sim 10^{-2}$ the corresponding multiplicity required is in order of $10^4$ particles, 
which is close to the parameters at the LHC. This is parametrically the same dependence as e.g. in a flow analysis in terms
of second order cumulants. Higher order cumulant expansions can improve this statistical limit. 
In this sense, the CMB-like analysis of heavy ion collisions shown here provides an alternative for characterizing collective flow
but does not offer obvious parametric advantages. However, it offers arguably a very well-suited setting for analyzing those
event-wise fluctuations that remain once the dominant azimuthal modulations have been removed. 
In analogy to the questions asked in cosmology, one can then characterize what sets  the scale of these remaining fluctuations. 
For instance, which physics underlies the fluctuations seen in a Mollweide projection of heavy ion collisions after removal of flow harmonics, 
such as in Fig.~\ref{fig_Mollweide_map}. How are these fluctuations 
sensitive to resonance decays, jet-like particle correlations or jets? On which scales could critical phenomena leave characteristic
signatures? Which similarities or characteristic differences could be expected for the corresponding maps based on fluctuations in
the distribution of electric charge, or strangeness content? We intend to follow up these and further questions in the analysis of 
real data, as well as in further model studies. Some of these questions are well known in CMB science, when we can use , for instance, optimal filter approach to amplify the point-like sources , incorporated into diffuse background. This method in combination with wavelet transform of the signal can potentially detect localized features with very small amplitude of heavy ion collisions, which can be associated with low amplitude jets. This work is in progress and will be published in a separate paper. 
\noindent
\section{Acknowledgements} We are grateful to  A.Bilandzic, S. Floerchinger, J. Nagle, R. Snellings and E. Shuryak
for very useful discussions , and anonymous referee for very stimulated comments.
We thank the Danish Natural Science Research Council (FNU) and the 
Danish National Research Foundation (Dansk Grundforsknings Fond) for support. O.V. thanks Dmitry Zimin's nonprofit Dynasty Foundation for the support.
This work is a part of CERN-LFI PLANCK collaboration. We are thankful to N. Mandolesi for support and 
useful discussions.

%%%%%%%%%%%%%%%%%%%%%%%%%%%%%%%%%%%%%%%%%%%%%%%%%%%%%%%%%%%%%%%%%%%%%%%%%%%%%
\begin{appendix}
\section{Stochastic equation and general description of  modulation}
In this appendix we discuss how the method of section IV can be generalized to an analysis that accounts for an arbitrary dependence of $v_n=v_n(\theta)$ on pseudo-rapditity 
$\eta=-\ln(\tan(\frac{\theta}{2}))$. This can be done by multiplying the stochastic equation by functions $W(\theta)$ with different shapes.
The stochastic equation for  modulations of the random signal is given by:
\begin{eqnarray}
 S(\theta,\phi)&=&f(\theta,\phi)\left[1+2\sum_nv_n\cos(n(\phi-\Psi_n))\right]=\nonumber\\
 f(\theta,\phi)&+&\sum_nv_n\left[e^{in\Psi_n}s^+(\theta,\phi)+e^{-in\Psi_n}s^-(\theta,\phi)\right],\nonumber\\
\label{eqa1}
\end{eqnarray}
where 
\begin{eqnarray}
 s^+(\theta,\phi)=f(\theta,\phi)e^{-in\phi},\hspace{0.1cm} s^-(\theta,\phi)=f(\theta,\phi)e^{in\phi}.
\label{eqa2}
\end{eqnarray}
The distribution $S(\theta,\phi)$ can have a significant $\theta$-dependence. We weigh the left and right hand side of Eq.(\ref{eqa1}) by the window function $W(\theta)$, 
\begin{eqnarray}
 &&\hat{F}S(\theta,\phi)W(\theta)=  \hat{F}S(\theta,\phi)W(\theta)+\nonumber\\
&&v_n \hat{F}W(\theta)\left[e^{in\Psi_n}s^+(\theta,\phi)+e^{-in\Psi_n}s^-(\theta,\phi)\right]\, ,
\label{eqa3}
\end{eqnarray}
where $\hat{F}$ is the linear operator of the spherical harmonic decomposition
\begin{eqnarray}
\hat{F} ... =\int_0^{2\pi}d\phi\int_0^{\pi}d\theta\sin\theta Y^*_{l,m}(\theta,\phi)...\, .
\label{eqa4}
\end{eqnarray}
In radioastronomy, this is referred to as the apodization approach and it serves to amplify
some particular range of the $(\theta,\phi)$ map, in order to reduce some very bright sources of contamination of the signal under investigation.
Here, we use it to select bins in $\theta$ for a characterization of the dependence of the signal on pseudo-rapidity. 
The coefficients of the spherical harmonic decomposition read then
\begin{eqnarray}
 b_{l,m}&=&c_{l,m}+v_n\int_0^{2\pi}\int_0^{\pi}d\theta\sin\theta\times\nonumber\\ &&Y^*_{l,m}(\theta,\phi)\left[e^{in\Psi_n}s^+(\theta,\phi)+e^{-in\Psi_n}s^-(\theta,\phi)\right].\nonumber\\
b_{l,m}&=&\hat{F}W(\theta)S(\theta,\phi),\hspace{0.1cm}c_{l,m}=\hat{F}W(\theta)f(\theta,\phi).
\label{eqa5}
\end{eqnarray}
Using the spherical harmonic decomposition $f(\theta,\phi)=\sum_{l',m'}a_{l',m'}Y_{l',m'}(\theta,\phi)$, from Eq(\ref{eqa5}) we have
\begin{eqnarray}
c_{l,m}&=&\hat{F}W(\theta)f(\theta,\phi)=\sum_{l',m'}a_{l',m'}G_{l,m}^{l',m'}(n=0),\nonumber\\
 s^+_{l,m}&=&\sum_{l',m'}a_{l',m'}G_{l,m}^{l',m'}(n)\, ,\nonumber\\
s^-_{l,m}&=&\sum_{l',m'}a_{l',m'}G_{l,m}^{l',m'}(-n)
\label{eqa6}
\end{eqnarray}
where
\begin{eqnarray}
G_{l,m}^{l',m'}(n)&=&\int_0^{2\pi}d\phi\int_0^{\pi}d\theta\sin\theta Y^*_{l,m}(\theta,\phi)\times\nonumber\\
&&Y_{l',m'}(\theta,\phi)W(\theta)e^{in\phi} \, .
\label{eqa7}
\end{eqnarray}
The coefficients of the spherical harmonic decomposition satisfy then
\begin{eqnarray}
 b_{l,m}=c_{l,m}+v_n\left[e^{in\Psi_n}s^+_{l,m}+e^{-in\Psi_n}s^-_{l,m}\right]\, .
\label{eqa8}
\end{eqnarray}
It is technically advantageous to introduce the variable $x=\cos\theta$. With the help of 
\begin{eqnarray}
 Y_{l,m}&=&\sqrt{\frac{(2l+1)}{4\pi}\frac{(l-m)!}{(l+m)!}}P^m_l(x)e^{im\phi}\nonumber \\
&=&N_{l,m}P^m_l(x)e^{im\phi}\, ,
\label{eqa9}
\end{eqnarray}
one can then express the coefficients in Eq.~(\ref{eqa8}) as
\begin{eqnarray}
 s^+_{l,m}&=&\sum_{l',m'}a_{l',m'}G_{l,m}^{l',m'}(n) \nonumber\\
&\simeq&\delta_{m,n}\sum_{l'} a_{l',0}G_{l,n}^{l',0}(n) +O(a_{lÂ´,0}),\nonumber\\
G_{l,n}^{l',0}(n)&=&2\pi N_{l,0}N_{l',n}\times\nonumber\\ &&\int_{-1}^1dx W(x)P^{0}_{l'}(x)P^{n}_{l}(x)
\label{eqa13}
\end{eqnarray}
The case of a $\theta$-independent distribution that we discussed predominantly in the main text,
is recovered for $W(x)=1$. In this case, one can use the integrals 
\begin{eqnarray}
\int_{-1}^1dx P_l^0(x)P_l^2(x)&=&-\frac{2l(l-1)}{2l+1},\nonumber\\
\int_{-1}^1dx\left[P^m_l(x)\right]^2&=&\frac{2}{2l+1}\frac{(l+m)!}{(l-m)!}
\label{eqa14}
\end{eqnarray}
to write
\begin{eqnarray}
 G_{2,2}^{2,0}(n=2)&=&-\frac{1}{\sqrt{6}},\hspace{0.5cm}G_{2,0}^{2,0}=1\, .
\label{eqa15}
\end{eqnarray}
According to  Eq.~(\ref{eq8}), this gives
\begin{eqnarray}
 v_2\simeq \sqrt{6}\frac{|b_{2,2}|}{|b_{2,0}|}.
\label{eqa16}
\end{eqnarray}

In order to analyze the $\theta $-dependency of the
amplitude of the flow $v_n(\theta)$, keeping in Eq.(\ref{eqa1}) the azimuthal modulations proportional to $\cos(n(\phi-\Psi_n))$.
Then, we can decompose the amplitude of the flow through Legendre polynomials:
\begin{eqnarray}
 v_n(\theta)=v_{n}\sum_qC_q^nP_q(\cos\theta)
\label{theta1}
\end{eqnarray}
where: $v_{n}=const$, and $C_q^n$ are the coefficients of decomposition.  
Obviously, the decomposition of $v_n(\theta)$ in the form of Eq.(\ref{theta1}) is not unique. One can use, for instance the associated Legendre
polynomials $P^n_l(\theta)$, instead of Legendre polynomials. The particular choice of the orthogonal functions should reflect the
properties of the effects of modulations under investigation. 

In the most general case, azimuthal modulations are no longer 
separable from polar ones. In heavy ion physics, this is the case e.g. for jet-like particle correlations. Instead of 
Eq.(\ref{eqa1}), the starting point for analyzing such modulations would be then the stochastic equation 
\begin{eqnarray}
 S(\theta,\phi)=f(\theta,\phi)\left[1+2V(\theta,\phi)\right]\, ,
\label{theta2}
\end{eqnarray}
where 
\begin{eqnarray}
 V(\theta,\phi)=\sum_l\sum_{m=-l}^lV_{l,m}Y_{l,m}(\theta,\phi)\, .
\label{theta3}
\end{eqnarray}

The coefficients $S_{l,m}$  of the spherical harmonic decomposition  can be expressed as $S_{l,m}=a_{l,m}+D_{l,m}$, where
  \begin{eqnarray}
  D_{l,m}&=&2 \sum_{{l'}=1}^{\infty}\sum_{m'=-l'}^{l'}\sum_{l''=0}^{\infty}\sum_{m''=-l''}^{l''}
  (-1)^{m} a_{l'm'}V_{l''m''}\times\nonumber\\
&\times&\sqrt{\frac{(2l''+1)(2l'+1)(2l+1)}{4\pi}}
  \left(\begin{array}{ccc}l'' & l' & l \\ 0 & 0 &0 \end{array}\right)\times\nonumber\\
 &\times& \left(\begin{array}{ccc}l'' & l' & l \\ m'' & m' & -m\end{array}\right)\, ,
 \label{lambdalm}
 \end{eqnarray}
and $a_{lm}=\hat F f(\theta,\phi)$.
This equation is well known in CMB physics (see for instance \citep{hansonlewis}). The major difference betwen modulation of
the signal in form of Eq.(\ref{theta3}) and $v_n=const$ is related to redistribution of the power from even multipoles to odd ones, 
if $V_{2n+1,m}$ has non-zero components. In the case when only $V_{2n,m}$-coefficients are non-vanishing, this quadrupole modulation will redistribute the power from the $n$-th mode to $n-1$,$n+1$ modes, where $l=2n$ and change the orientation of the quadrupole from very planar ( $v_2=const$) to non-planar, depending on amplitude of $2,1$-component. We shall explore in future work 
to what extent Eqs.~(\ref{theta3}) and (\ref{lambdalm}) allow to investigate more complex modulations of the particle distribution in high-multiplicity heavy ion collisions. 

\end{appendix}

\end{document}